\documentclass[acmsmall,nonacm]{acmart}

\usepackage{graphicx} 
\usepackage{subcaption} 
\usepackage{booktabs,tabularx}
\usepackage{ifthen}
\usepackage{tabularx}
\usepackage{colortbl}
\usepackage{amsmath}
\usepackage{amsthm}
\usepackage{mathtools}
\usepackage{amsfonts}
\usepackage{wrapfig}
\usepackage{paralist}
\usepackage{quantikz}
\usepackage{xspace}
\usepackage{trimclip}   \usepackage[capitalise]{cleveref}
\usepackage{tikz}
\usepackage{stmaryrd}
\usepackage{todonotes}
\usepackage{multicol}
\usepackage{forest}
\usepackage[ruled, vlined, linesnumbered]{algorithm2e}
\usepackage{multirow}
\usepackage{thmtools}
\usepackage{thm-restate}
\usepackage{soul}
\usepackage{framed}
\usepackage{braket}
\usetikzlibrary{arrows}
\usetikzlibrary{positioning}
\usetikzlibrary{cd}
\usetikzlibrary{patterns.meta}
\usepackage{microtype}
\usepackage{adjustbox}
\usepackage{wrapfig}

\renewcommand{\i}[0]{\mathrm{i}}
\DeclareMathOperator*{\tr}{tr}
\newcommand{\anc}[0]{\mathrm{anc}}
\newcommand{\sys}[0]{\mathrm{sys}}

\newcommand{\ii}{\mathrm{i}}
\newcommand{\Sabbr}{\mathcal{S}}
\newcommand{\Tabbr}{\mathcal{T}}

\usepackage{transparent}
\usetikzlibrary {graphs,quotes}

\usepackage{tikz}
\usetikzlibrary{quantikz2}

\usepackage{booktabs}
\usepackage{array}

\newcommand{\showcomment}[1]{}

\newcommand{\rmark}[1]{{#1 }}

\renewcommand{\varepsilon}{\epsilon}

\usepackage{pgfplots}
\usepackage[mode=tex]{standalone}

\newcommand{\algfont}[1]{\textnormal{\textsc{#1}}}
\newcommand{\Accept}[0]{\algfont{Accept}}
\newcommand{\Reject}[0]{\algfont{Reject}}
\newcommand{\VerifyBlockEncoding}[0]{\algfont{VerifyBlockEncoding}}
\newcommand{\SymbolicExecute}[0]{\algfont{SymbolicExecute}}
\newcommand{\ExtractPolynomial}[0]{\algfont{ExtractPolynomial}}
\newcommand{\VerifyPolynomialApproximation}[0]{\algfont{VerifyPolynomialApproximation}}
\newcommand{\VerifyHamiltonianSimulation}[0]{\algfont{VerifyHamiltonianSimulation}}
\newcommand{\QSP}[0]{\algfont{QSP}}
\newcommand{\BE}[0]{\algfont{BE}}

\newcommand{\I}[0]{\mathcal{I}}

\newcommand{\eq}[1]{(\hyperref[eq:#1]{\ref*{eq:#1}})}
\renewcommand{\sec}[1]{Section~\hyperref[sec:#1]{\ref*{sec:#1}}}
\newcommand{\thm}[1]{Theorem~\hyperref[thm:#1]{\ref*{thm:#1}}}
\newcommand{\lem}[1]{Lemma~\hyperref[lem:#1]{\ref*{lem:#1}}}
\newcommand{\cor}[1]{Corollary~\hyperref[cor:#1]{\ref*{cor:#1}}}
\newcommand{\itm}[1]{Item~\hyperref[itm:#1]{\ref*{itm:#1}}}
\newcommand{\app}[1]{Appendix~\hyperref[app:#1]{\ref*{app:#1}}}
\newcommand{\dfn}[1]{Definition~\hyperref[dfn:#1]{\ref*{dfn:#1}}}
\newcommand{\fig}[1]{Fig.~\hyperref[fig:#1]{\ref*{fig:#1}}}
\newcommand{\alg}[1]{Algorithm~\hyperref[alg:#1]{\ref*{alg:#1}}}
\newcommand{\asm}[1]{Assumption~\hyperref[asm:#1]{\ref*{asm:#1}}}
\newcommand{\prot}[1]{Protocol~\hyperref[prot:#1]{\ref*{prot:#1}}}
\newcommand{\prob}[1]{Problem~\hyperref[prob:#1]{\ref*{prob:#1}}}
\newcommand{\rmk}[1]{Remark~\hyperref[rmk:#1]{\ref*{rmk:#1}}}
\newcommand{\tabl}[1]{Table~\hyperref[tabl:#1]{\ref*{tabl:#1}}}
\newcommand{\prop}[1]{Proposition~\hyperref[prop:#1]{\ref*{prop:#1}}}

\begin{document}
\title{Automated Verification of Hamiltonian Simulation by Quantum Signal Processing}
\author{Min Lai}
\affiliation{\institution{National Taiwan University}\country{Taiwan}}
\email{mimi.lai.cat@gmail.com}
\author{Yu-Fang Chen}
\affiliation{\institution{Academia Sinica}\country{Taiwan}}
\email{yfc@iis.sinica.edu.tw}
\author{Shih-Han Hung}
\affiliation{\institution{National Taiwan University}\country{Taiwan}}
\email{shihhanh@ntu.edu.tw}
\begin{abstract}
We present, to the best of our knowledge, the first fully automated
verification framework for QSP-based Hamiltonian simulation. Our framework
exploits the structure of the simulation algorithm: Hamiltonians are given as weighted
sums of Pauli strings, embedded into larger unitaries through block encodings,
and transformed into time-evolution operators by quantum signal processing (QSP).
Using a symbolic Pauli-sum semantics for ancilla-indexed circuit blocks, our
verifier proves that the distinguished all-zero ancilla block is proportional
to the target Hamiltonian and infers the subnormalization factor automatically.
It then verifies the QSP layer by extracting the implemented polynomial
transformation and certifying its approximation error.
This structure-aware abstraction is essential for scalability. 
Composing our verified block encoding and QSP wrapper yields
an estimated macro-expanded size of about 1.52 million gates.
The verifier certifies the block encoding once and checks the QSP layer
through its abstract interface, without expanding the oracle calls. Our implementation verifies block encodings of structured Pauli-sum Hamiltonians with up to 256 system qubits and verifies QSP Hamiltonian-simulation instances by extracting
their implemented polynomials and checking the resulting approximation bounds.
These results show that fully automated verification can handle common
Hamiltonian-simulation circuits at substantial scale when it exploits their algorithmic structure.
\end{abstract}
\maketitle

\section{Introduction}\label{sec:intro}
Hamiltonian simulation is one of the central algorithmic tasks in quantum computing and a leading application area where quantum computers may provide computational advantage. It underlies applications in quantum chemistry, condensed-matter physics, and quantum algorithms more broadly. 
The task is defined as follows: given a Hamiltonian $H$ and time $t$, the goal is to approximate the time-evolution operator $e^{-\i H t}$ to within error $\epsilon$ \cite{doi:10.1126/science.273.5278.1073}. 

Since \citet{feynmanSimulatingPhysicsComputers1982} proposed the idea of simulating quantum systems using a quantum computer and \citet{doi:10.1126/science.273.5278.1073} gave the first explicit algorithm, many quantum algorithms have been given to tackle the problem. 
In particular, three families of quantum algorithms dominate. 
\emph{Product formulas} approximate $e^{-\i H t}$ by interleaving the evolutions under the individual terms of $H=\sum_j H_j$, and their correctness rests on nested commutator bounds \cite{PhysRevX.11.011020}. 
\emph{Linear combination of unitaries} (LCU) approximates $e^{-\i H t}$ as a weighted sum of easy unitaries via an ancilla-controlled selection circuit \cite{childsHamiltonianSimulationUsing2012}. 
\emph{Quantum signal processing} (QSP) achieves nearly optimal complexity by encoding a polynomial approximation of $e^{-\i x t}$ into the eigenphases of a block-encoding walk operator \cite{PhysRevLett.118.010501,Low2019hamiltonian}.

The last two families share a common primitive: a \emph{block encoding} of matrix $H$, a unitary that embeds the subnormalized operator $H/\alpha$ as a submatrix (in standard form, its top-left block) for some subnormalization factor $\alpha$.
This allows non-unitary $H$ to be accessed coherently on an enlarged quantum state. 
A defining feature of these algorithms is that no implementation ever \emph{equals} the target $e^{-\i H t}$: the algorithms are deliberate truncations governed by a tunable accuracy parameter, and a finite universal gate set can in any case only approximate the required rotations, as analyzed by \citet{2011679.2011685}. 
The obligation of the verification task is therefore not just exact equivalence checking, but an \emph{approximate} generalization of it, i.e., deciding whether two quantum processes agree to within error $\epsilon$ in a suitable metric.

This is substantially harder than, and poorly served by the tools built for, its exact counterpart. 
The difficulty is not merely numerical.
The operators are $2^n$-dimensional and given only implicitly by a circuit, so approximate checking must certify a quantitative norm bound on an object it cannot represent explicitly. 
Furthermore, the existing techniques that make exact equivalence tractable---decision diagrams \cite{burgholzer2020advanced}, stabilizer formalism \cite{ThanosCL23}, path-sums \cite{amy2018towards}, weighted model-counting \cite{huang2026equivalence}, tree automata~\cite{ChenCLLT23,ChenCLLTY23,chen2025autoq}, and ZX-calculus rewriting \cite{Coecke_2011}---primarily target circuit-to-circuit comparisons or state-based properties, rather than directly certifying a spectral-norm bound against an operator-level specification without a reference circuit. 

\paragraph{Automated verification of Hamiltonian simulation}
Our main contribution is a novel solution to automatically verifying Hamiltonian simulation based on quantum signal processing. 
Here, Hamiltonian simulation verification means proving that a quantum circuit generated from a Hamiltonian $H$ correctly approximates the intended time-evolution operator $e^{-\i Ht}$. 

This gap is especially visible in \emph{block-encoding-based} Hamiltonian simulation. In these algorithms, correctness is not a statement that the whole circuit unitary equals another circuit. The Hamiltonian is typically given
symbolically as a weighted sum of Pauli strings, and the verifier must first show that an ancilla-selected block of a larger unitary is proportional to this Hamiltonian, while also recovering the subnormalization factor.

A second semantic obligation arises from the simulation layer built around the
certified block encoding. In QSP-based Hamiltonian simulation, the circuit
repeatedly invokes the block encoding and its adjoint, interleaved with phase
transformations, so that the overall circuit block-encodes a polynomial transformation of the encoded Hamiltonian. The final check is therefore an
approximation guarantee: after accounting for the block-encoding subnormalization,
the implemented polynomial must approximate $e^{-\i Ht}$ within the prescribed
error bound. Thus, verification naturally decomposes into certifying the block encoding and certifying the QSP transformation.

Fully automated verification of this kind remains largely out of reach for existing general-purpose quantum verification techniques. The difficulty is not merely circuit size, but the semantic gap between the generated circuit and the high-level specification it is meant to satisfy. A verifier must connect low-level gates to the high-level specification
$e^{-\i Ht}$. This requires certifying the block encoding of the symbolic
Pauli-sum Hamiltonian, reasoning about ancilla-selected blocks rather than
entire unitaries, and verifying that the surrounding QSP layer implements the
intended polynomial approximation of time evolution. 

We present, to the best of our knowledge, the first fully automated
verification framework for block-encoding-based Hamiltonian simulation
circuits. Our verifier combines two complementary symbolic techniques. For
block encodings, it constructs a branch-indexed symbolic model of the circuit
by symbolic execution, where each branch operator is represented as a weighted
sum of Pauli strings. This allows the verifier to inspect the induced
all-zero ancilla block, prove that it is proportional to the target
Hamiltonian, and infer the subnormalization factor automatically.

The resulting block-encoding certificate is then used as an abstract interface
for verifying the simulation layer. In this paper, we focus on QSP-based
Hamiltonian simulation. For the QSP component, the verifier replaces each
invocation of the certified block encoding by a four-block symbolic
abstraction, symbolically executes the QSP circuit, and uses sound rewrite
rules derived from unitarity to extract the implemented polynomial
transformation. The extracted polynomial is then checked against the desired
time-evolution function after accounting for the block-encoding subnormalization.
Together, these two symbolic components validate the two main obligations in
QSP-based Hamiltonian simulation: certifying the input block encoding and
certifying the surrounding QSP transformation.
Although QSP is our main target, the same decomposition also suggests a path
toward verifying other block-encoding-based simulation methods, including
LCU-based algorithms derived from truncated Taylor-series approximations.

Our evaluation shows that this structure-aware approach scales to substantial
Hamiltonian-simulation instances. The verifier checks block encodings of
structured Pauli-sum Hamiltonians with up to 256 system qubits and hundreds of
Pauli terms, without constructing the exponentially large system matrix. It
also verifies QSP-based Hamiltonian-simulation circuits by extracting the
implemented polynomial and certifying its approximation error.
Composing the verified block encodings and QSP wrappers yields large
Hamiltonian-simulation circuits whose correctness can be established
without expanding the oracle calls. Our evaluation illustrates this
composition and reports the resulting error bound and macro-expanded
circuit size. These results demonstrate fully automated
verification of common block-encoding-based Hamiltonian-simulation circuits at
substantial scale.
\section{Background}

\subsection{Quantum Computing}

Here we give a very brief introduction to quantum computing and fix notation. For a full treatment, we refer the readers to \citet{Nielsen_Chuang_2010}.

\paragraph{States.} The state of a single qubit is a unit vector in the Hilbert space $\mathbb C^{2}$, written in the computational basis as a \emph{superposition} $\ket{\psi}=\alpha\ket{0}+\beta\ket{1}$ with $|\alpha|^2+|\beta|^2=1$.
While a single-qubit state is represented using two complex numbers, only the magnitudes and the relative phase of the coefficients $\alpha$ and $\beta$ matter in algorithmic settings, since the global phase is physically irrelevant. 
An $n$-qubit state lives in the tensor product $(\mathbb C^2)^{\otimes n}\cong \mathbb C^{2^n}$.
The exponential dimension is the source of quantum states' expressive power and of the difficulty of classical simulation algorithms. 

\paragraph{Evolution.} Elementary operations that describe the dynamics of quantum systems are unitaries: a computation transforms $\ket{\psi}$ to $U\ket{\psi}$ for some unitary $U$, satisfying $U^\dag U=UU^\dag =I$. 
Algorithms are expressed as \emph{quantum circuits} built from a universal gate set, which uniformly approximates any unitary to arbitrary accuracy. 
A unitary may be written as the time evolution in the form $e^{-\i H t}$ for some Hamiltonian---a Hermitian matrix $H$ of the same size. 
As we will be discussing in this paper, we can implement a unitary $U$ when the corresponding $H$ can be written as a linear combination of unitaries with few terms. 

We will be frequently talking about the Pauli matrices, which form a set of unitary matrices closed under matrix multiplication up to a phase in $\{1,\i,-1,-\i\}$. 
A single-qubit Pauli matrices include four operators, in standard basis written as:
\begin{align}
    I = \begin{bmatrix}
        1 & 0 \\ 0 & 1 
    \end{bmatrix},
    \quad
    X = \begin{bmatrix}
        0 & 1 \\ 1 & 0
    \end{bmatrix},
    \quad
    Y = \begin{bmatrix}
        0 & -\i \\ \i & 0
    \end{bmatrix},
    \quad
    Z = \begin{bmatrix}
        1 & 0 \\ 0 & -1
    \end{bmatrix}
\end{align}
It is straightforward to check the multiplication of any two matrices yields another up to a phase. 
An $n$-qubit Pauli matrix is the tensor product of single-qubit Pauli matrices, i.e., they form a set $\mathcal P_n=\{I,X,Y,Z\}^{\otimes n}$.
The Pauli matrices are orthogonal to one another with respect to the Hilbert-Schmidt inner product---that is, for $P,Q\in\mathcal P_n$, $\langle P,Q\rangle = \tr(P^\dag Q)=0$ if and only if $P\neq Q$. 
Furthermore, they form a complete basis for matrices of the same size. 
As a result, we can uniquely decompose any matrix into a linear combination of Pauli matrices. 
Moreover, the coefficients are real when the matrix is Hermitian.

\paragraph{Measurement.}
Information is extracted by measurement. 
A projective measurement of a state $\ket{\psi}$ in the computational basis yields an outcome $x$ with probability $|\langle x|\psi\rangle|^2$ (Born's rule). 
Upon measurement, the quantum state collapses to the outcome state $\ket{x}$.
Measurement is thus probabilistic. 

\paragraph{Norms and spectra.}
For a complex number $z\in\mathbb C$, we write $|z|$ for its modulus.
For a vector $v\in\mathbb C^n$, we write $\|v\|$ for the Euclidean norm,
and for vectors $u,v\in\mathbb C^n$ we write
$\langle u,v\rangle=\sum_j \overline{u_j}v_j$ for the standard complex inner product, where $\overline{u_j}$ is the complex conjugate of $u_j$.  For a matrix $A$, we write
$\|A\|$ for its spectral norm, defined by
\[
    \|A\|=\max_{\|v\|=1}\|Av\| .
\]
Equivalently, $\|A\|$ is the largest singular value of $A$; when $A$ is normal,
it is the largest absolute value of its eigenvalues.  We write $\|A\|_F$ for
the Frobenius norm,
\[
    \|A\|_F=\sqrt{\tr(A^\dagger A)}.
\]
Equivalently, if $\sigma_1(A),\ldots,\sigma_r(A)$ are the singular values of
$A$, then $\|A\|_F=\sqrt{\sum_{j=1}^r \sigma_j(A)^2}$.
For a Hermitian matrix $H$, we write $\mathrm{Spec}(H)$ for the set of
eigenvalues of $H$.  Since Hermitian matrices have real eigenvalues,
$\|H\|\le 1$ implies $\mathrm{Spec}(H)\subseteq[-1,1]$.

\subsection{Block Encoding in Quantum Computational Linear Algebra}

Block encoding of an exponentially large linear operator is an essential subroutine for the design of a wide range of quantum algorithms \cite{PRXQuantum.2.040203}.
A unitary operator $U$ block encodes a linear operator $A$ if the topleft diagonal submatrix is a scalar multiple of $A$. 
While the matrix $A$ to be block encoded may be a non-square matrix in general, we restrict our discussion to square matrices, since we will be focusing on Hamiltonian simulation.

\begin{definition}[Block encoding]
    A unitary $U\in\mathrm U(2^{n+m})$ block encodes a matrix $A\in\mathbb C^{2^n\times 2^n}$ using $m$ ancilla qubits if 
    \begin{align}
        U = \begin{bmatrix}
            A/\alpha & \cdot \\ \cdot & \cdot
        \end{bmatrix} \iff
        A/\alpha = (\bra{0^m}\otimes I^{\otimes n}) U (\ket{0^m}\otimes I^{\otimes n}),
    \end{align}
    where $I$ is an $2\times 2$ identity matrix.
    The fact that the topleft submatrix scaled by a real number $\alpha\geq\|A\|$ follows from the fact that $U$ is a unitary matrix. 
    A quantum circuit block encodes $A$ if it produces a unitary operator that block encodes $A$.
    A block encoding of $A$ will often be denoted as $U_A$, when the implementation detail is irrelevant in context.
\end{definition}

\paragraph{Linear Combination of Unitaries.}
A standard way to construct block encodings is the Linear Combination of Unitaries
(LCU) method. 
Suppose $A=\sum_{i=0}^{M-1} c_iU_i$ is a linear combination of $M$ unitaries.
Let $\alpha=\sum_i |c_i|$, and set $m=\lceil\log M\rceil$ after padding the
list to length $2^m$ if necessary. The unitary PREPARE maps
$\ket{0^m}$ to $\sum_{i=0}^{2^m-1}\sqrt{|c_i|/\alpha}\ket{i}$, and the
SELECT unitary applies the phase-weighted operation $(c_i/|c_i|)U_i$ when the
ancilla is in state $\ket{i}$, with an arbitrary phase convention when
$c_i=0$. Then
\[
    ( \bra{0^m}\otimes I)(\mathrm{PREPARE}^\dagger\otimes I)\mathrm{SELECT}(\mathrm{PREPARE}\otimes I)
    ( \ket{0^m}\otimes I)
    =
    A/\alpha.
\]
Thus $(\mathrm{PREPARE}^\dagger\otimes I)\mathrm{SELECT}(\mathrm{PREPARE}\otimes I)$ block encodes $A$ with
subnormalization factor $\alpha$. The same construction also applies when the
controlled operations are themselves block encodings, yielding a block encoding
of the corresponding weighted sum. We next illustrate LCU with a simple block
encoding of a sum of two single-qubit Pauli operators.

\begin{example}\label{ex:lcu}
    We can use the quantum circuit from Fig.~\ref{fig:example-block-encoding-circuit} to block encode the operator $X-Z$: 
    \begin{figure}[hbtp]
    \begin{center}
    \begin{quantikz}
        \lstick{$\ket{0}_{\anc}$} & \gate{X} & \gate{H} & \ctrl[open]{1} & \ctrl{1} & \gate{H} & \meter{} & \setwiretype{c} \rstick{0} \\
        \lstick{$\ket{\psi}_{\sys}$} & \qw & \qw & \gate{X} & \gate{Z} & \qw & \qw & \rstick{$\frac{X-Z}{2}\ket{\psi}_{\sys}$}
    \end{quantikz}
    \end{center}
    \caption{A block encoding of $X-Z$ with a subnormalization factor $2$.}
    \label{fig:example-block-encoding-circuit}
    \end{figure}
    Upon measurement with outcome $0$, the state of the remaining wire is $\frac{X-Z}{2}\ket{\psi}$.
    Therefore, this circuit block encodes $X-Z$ with subnormalization factor $2$ using one ancilla qubit, i.e., we have
    \begin{align}
        \left(\bra{0}_{\anc}\otimes I_{\sys}\right) C \left(\ket{0}_{\anc}\otimes I_{\sys}\right) = \frac{X-Z}{2}.
    \end{align}      
\end{example}
To emphasize the roles of the qubits, we will denote the qubit that starts with a zero state and ends with a zero outcome as the ancilla qubits, and the qubits whose state is unrestricted are called the system qubits.

\paragraph{Pauli Decomposition} 
Recall that the Pauli matrices $\{I,X,Y,Z\}^{\otimes n}$ form an orthogonal basis with respect to the Hilbert--Schmidt inner product.
We can uniquely decompose a matrix of the same size into a linear combination of the Pauli matrices. 
Therefore given any $2^n\times 2^n$ matrix $A$, we can compute a block encoding by its decomposition into Pauli matrices of the same dimension, and apply LCU to implement a block encoding. 
While this method applies generally to any matrix, we should note that it does not produce an \emph{efficient} block encoding, unless the number of terms in the summation is polynomially bounded.

\subsection{Quantum Signal Processing}

With the ability to perform matrix arithmetic using block encodings, we can further construct functions of encoded matrices. 
Quantum Singular Value Transformation (QSVT), formally established by \citet{10.1145/3313276.3316366}, provides a systematic way of producing block encodings of a polynomial $p$ applied to the singular values of an encoded matrix $A$. 
When $A$ is Hermitian, as in the Hamiltonian simulation setting discussed below, this framework effectively evaluates $p(A)$.

Given a block encoding $U_A$ of Hermitian $A$ and some target degree-$d$ polynomial $p$, an ancilla-efficient method to evaluate $p(A)$ is provided by the framework of Quantum Signal Processing (QSP).
For a polynomial function $p$ satisfying the standard QSP admissibility conditions, including the required parity and boundedness constraints, there is a sequence of angles $\phi_0,\ldots,\phi_d$ such that the following sequence of unitaries produces a block encoding whose real part is $p(A)$:
\begin{align}\label{eq:qsp-circuit}
    U_{p(A)} = e^{\i\phi_d Z_{\Pi}} (U_A)^\dag e^{\i\phi_{d-1} Z_{\Pi}} (U_A) \ldots (U_A)^\dag e^{\i\phi_{1} Z_{\Pi}} (U_A) e^{\i\phi_0 Z_{\Pi}},
\end{align}
where $Z_\Pi$ is a \emph{reflection} associated with the projector $\Pi$, i.e., $Z_\Pi = 2\Pi-I$ \cite{10.1145/3313276.3316366}.
Furthermore, given the coefficients of $f$, the corresponding angles can be efficiently found using a classical computer \cite{10.1145/3313276.3316366}.

Although QSP directly implements polynomial transformations, it can also be used to approximate non-polynomial functions that admit accurate polynomial approximations on the relevant spectral interval.
More precisely, if a non-polynomial function can be uniformly approximated by a polynomial of sufficiently large degree over the singular values of interest, then QSP can be used to implement the corresponding approximate transformation.
In the context of Hamiltonian simulation, to approximate a block encoding of the time evolution operator $e^{-\i Ht}$ for a Hamiltonian $H$---a Hermitian matrix describing a physical system---one can approximate the scalar function $e^{-\i x t}$ by a polynomial of sufficiently high degree, and then apply the QSP framework to obtain an approximation of $e^{-\i H t}$ \cite{PhysRevLett.118.010501}. 

\begin{example}[Implementation of QSP circuits]\label{ex:cosine-truncation}
    For Hamiltonian $H$ satisfying $\|H\|\leq 1$, we can approximate the real part of $e^{-\i H t}$, i.e., $\cos(Ht)$, by applying a truncation to the Jacobi-Anger expansion: 
    $p(H) = J_0(t) - 2 J_2(t) T_2(H) + 2 J_4(t) T_4(H)$,
    where $T_2(x)=2x^2-1$ and $T_4(x)=8x^4-8x^2+1$ are Chebyshev polynomials, and $J_m$ is the $m$-th Bessel function of the first kind.
    For a small evolution time $t=2$, we can numerically compute the coefficients:
\begin{align} \nonumber \cos(2x) 
    &\approx J_0(2) - 2 J_2(2) T_2(x) + 2J_4(2) T_4(x) \\\nonumber
    &= 0.22389078 - 0.70566806T_2(x) + 0.06799144T_4(x) \\\label{eq:approximate-p-high-precision}
    &= 0.54393152 x^4 - 1.95526763 x^2 + 0.99755028.
\end{align}
    We may compute the angles 
\begin{equation}
\phi_0\approx2.30473309,\quad
\phi_1=\phi_3\approx-1.05394248,\quad
\phi_2\approx1.06662014,\quad
\phi_4\approx0.73393676.
\label{eq:angles}
\end{equation}    
    such that (\ref{eq:qsp-circuit}) is satisfied.
    \qed
\end{example}

\section{Overview of the Verification Framework}

We now give an overview of our verification framework for
block-encoding-based Hamiltonian simulation. The goal is to verify that a
circuit generated from a symbolic Hamiltonian $H$ and simulation time $t$
implements an approximation of the time-evolution operator $e^{-iHt}$ within
a prescribed error bound $\varepsilon$. Rather than treating the whole circuit as an
unstructured unitary, the verifier follows the algorithmic structure of the
construction: it first certifies the block encoding, then symbolically analyzes
the QSP layer, and finally checks the extracted polynomial approximation.

Given a Hamiltonian $H$ written as a sum of Pauli operators, the verification
task proceeds in the following four steps.

\begin{wraptable}{r}{0.3\textwidth}
\vspace{-1.2em}
\centering
\small
\begin{tabular}{l|cc}
\toprule
$op$ & $B_0$ & $B_1$ \\
\midrule
Initial & $I$ & $0$ \\
$X(q_a)$ & $0$ & $I$ \\
$H(q_a)$ & $\frac{I}{\sqrt{2}}$ & $-\frac{I}{\sqrt{2}}$ \\
$\mathrm{open}\text{-}CX(q_a,q_s)$
& $\frac{X}{\sqrt{2}}$
& $-\frac{I}{\sqrt{2}}$ \\
$CZ(q_a,q_s)$
& $\frac{X}{\sqrt{2}}$
& $-\frac{Z}{\sqrt{2}}$ \\
$H(q_a)$
& $\mathbf{\frac{X-Z}{2}}$
& $\frac{X+Z}{2}$ \\
\bottomrule
\end{tabular}
\caption{Symbolic execution of the block-encoding circuit in
Example~\ref{ex:lcu}. Here $\mathrm{open}\text{-}CX$ is activated on ancilla
value $0$.}
\label{tab:example-block-encoding-symbolic-execution}
\vspace{-1.2em}
\end{wraptable}

\paragraph{\bfseries Verification of block encoding.}
The input to this step is a symbolic Hamiltonian
$H=\sum_{i=1}^M \alpha_iP_i$, given as Pauli strings $P_i$ with weights
$\alpha_i$, together with a circuit $Q$ alleged to block encode $H$. The verifier
checks whether $Q$ implements a block encoding of $H$ up to an unknown
subnormalization factor. Rather than constructing the full unitary matrix of
$Q$, it symbolically executes the circuit and tracks the system operator induced
on each ancilla branch.

\begin{example}
Consider the single-qubit Hamiltonian $H=(X-Z)/2$ and the LCU-style
block-encoding circuit $Q_{\mathrm{BE}}$ from Example~\ref{ex:lcu}, using one
ancilla qubit $q_a$ and one system qubit $q_s$. Since there is one ancilla qubit,
symbolic execution maintains two branch operators $B_0$ and $B_1$, representing
\[
    \ket{\Phi}
    =
    \ket{0}_{\mathsf{anc}}B_0\ket{\psi}_{\mathsf{sys}}
    +
    \ket{1}_{\mathsf{anc}}B_1\ket{\psi}_{\mathsf{sys}}.
\]
Initially $B_0=I$ and $B_1=0$. The verifier then updates these symbolic
operators gate by gate, as shown in
Table~\ref{tab:example-block-encoding-symbolic-execution}. For example,
$X(q_a)$ swaps the two branches, so $(B_0,B_1)=(I,0)\mapsto(0,I)$, while
$\mathrm{open}\text{-}CX(q_a,q_s)$ applies $X$ only on the branch where the
ancilla value is $0$, updating $B_0\mapsto XB_0$ and leaving $B_1$ unchanged.
After the final gate, the all-zero branch is $B_0=(X-Z)/2=H$. Therefore, the
verifier accepts $Q_{\mathrm{BE}}$ as a valid block encoding of $H$. More
generally, if the final all-zero branch operator is $H/\alpha$, then the circuit
block-encodes $H$ with subnormalization factor $\alpha$.
\qed
\end{example}
    \paragraph{\bfseries Symbolic execution of QSP} The input to this procedure is the Hamiltonian-simulation QSP circuit $C_{\mathrm{QSP}}$, together with the verified block-encoding gate
    $Q_{\mathrm{BE}}$. After the block-encoding circuit has been verified, the
    verifier no longer expands its internal gate-level implementation. Instead, the
    verified block-encoding circuit is treated as a symbolic unitary
    \[
    [\![Q_{\mathrm{BE}}]\!]
    =
    \begin{bmatrix}
    H & G \\
    A & C
    \end{bmatrix},
    \]
    where the top-left block $H$ is the verified encoded Hamiltonian, while
    $G,A,C$ are kept as symbolic blocks.
    In general, after the circuit $Q_{\mathrm{BE}}$ is certified, $Q_{\mathrm{BE}}$ is a block encoding of a $1/\alpha$ multiple of the target Hamiltonian. 
    To simulate the target Hamiltonian at evolution time $t$, effectively we can simulate $H$---the top-left block of $[\![Q_{\mathrm{BE}}]\!]$---at time $\alpha t$.
Therefore, for ease of notation, in the remaining three stages, we will follow the convention. 
    Namely, we simulate a Hamiltonian $H$ satisfying that $\|H\|\leq 1$ at evolution time $\alpha t$, when $H$ is the top-left block of $[\![Q_{\mathrm{BE}}]\!]$, and after $\alpha$ is obtained from the previous stage.
    The verifier symbolically executes $C_{\mathrm{QSP}}$ over this abstract
    block-encoding gate. Since the verifier reasons only about the principal block and the aggregate non-principal block, it supports only QSP circuits whose operations preserve this abstraction. In particular, the block-encoding ancilla may only be accessed through operations that distinguish the all-zero ancilla state from all remaining ancilla states collectively, while the system register is modified only through invocations of the certified block encoding.
    The output of this procedure is the operator encoded in the
    all-zero ancilla branch.

    \begin{example}
    Consider a Hamiltonian-simulation QSP circuit $C_{\mathrm{QSP}}$ together with a verified
    block encoding $Q_{\mathrm{BE}}$. Unlike the block-encoding verifier, which explicitly tracks every
    block-encoding ancilla branch, the QSP verifier reasons only about whether the block-encoding ancilla is in the all-zero state or not. Consequently, the internal implementation of the certified block encoding is no longer expanded. Every invocation of $Q_{\mathrm{BE}}$ is replaced by the symbolic four-block abstraction.

    The QSP verifier assumes a restricted circuit language.
    Between two consecutive calls to
    $Q_{\mathrm{BE}}$ or $Q_{\mathrm{BE}}^\dagger$,
    no gate modifies the system register.
    Moreover, operations on the block-encoding ancilla may depend only on whether the ancilla is in the all-zero state, and do not distinguish individual nonzero ancilla states.
    Consequently, every symbolic operator generated during execution is constructed solely from the symbolic blocks $H$, $A$, $G$, and $C$, together with scalar phase factors.
 
    On each call to the certified block encoding, the verifier applies the
    symbolic update
    
    \[
    \begin{bmatrix}
    B_0'\\
    B'_{\neq 0}
    \end{bmatrix}
    =
    \begin{bmatrix}
    H&G\\
    A&C
    \end{bmatrix}
    \begin{bmatrix}
    B_0\\
    B_{\neq 0}
    \end{bmatrix},
    \]
    where $B_0$ denotes the symbolic operator associated with the principal
    (all-zero) block-encoding branch, and $B_{\neq 0}$ collectively represents  all remaining block-encoding branches, rather than tracking them individually.
    Thus the symbolic expressions produced during execution may contain the auxiliary symbolic blocks $A$, $G$, and $C$ in addition to $H$.
\qed
\end{example}

\paragraph{\bfseries Polynomial Extraction by Unitary Simplification}
After symbolic execution finishes, the principal branch $B_0$ is an expression over $H$, $A$, $G$, and $C$. The verifier then applies a fixed set of oriented rewrite rules, derived from the unitarity identities of the four-block abstraction, to normalize this expression. These rewrite rules eliminate the auxiliary symbolic blocks whenever possible, and the verifier accepts only when the final expression reduces to a polynomial in $H$.

\begin{example}
Suppose symbolic execution produces
$E=HA^\dagger A+GG^\dagger H$. Using the unitarity identities
$A^\dagger A=I-H^\dagger H$ and $GG^\dagger=I-HH^\dagger$, the verifier rewrites
\[
E
=H(I-H^\dagger H)+(I-HH^\dagger)H 
=2H-HH^\dagger H-HH^\dagger H.
\]
In the Hamiltonian simulation setting, the encoded Hamiltonian is Hermitian, so
$H^\dagger=H$, and hence $E=2H-2H^3$. Thus the auxiliary blocks have been
eliminated, and the verifier reads the result as the polynomial
$f(x)=2x-2x^3$.
\end{example}

    \paragraph{\bfseries 
    Verification of approximate polynomial} 
    In the last step, the verifier checks whether the extracted polynomial
    approximates the desired time-evolution function. If the verified block
    encoding has principal block $H/\alpha$, then the QSP circuit applies the
    extracted polynomial $p$ to the normalized Hamiltonian $H/\alpha$. Therefore, to
    simulate the original Hamiltonian $H$ for time $t$, it suffices to check the
    scalar approximation
    \[
        p(x) \approx e^{-i\alpha x t}/\beta
        =
        e^{-i x(\alpha t)}/\beta
        \qquad
        \text{for } x\in[-1,1].
    \]
    for some complex subnormalization factor $\beta$.
    Thus, the block-encoding subnormalization $\alpha$ is compensated by simulating at an \emph{effective evolution
    time} $\alpha t$.
    
    The new  subnormalization factor $\beta$ may differ from 1 if the implementation chooses the QSP angles to compute a polynomial $p$ that approximates a scalar multiple of the exponential function. 
    In known implementations, e.g., \citet{10.1145/3313276.3316366}, the QSP step yields some subnormalization factors $|\beta|>1$ and compensates it by applying an oblivious amplitude amplification on top. 
    Therefore, to offer flexibility, our verification allows for $|\beta|\geq 1$ and computes it in the runtime.
    More specifically, the verification computes $\beta$, 
    and aims to check if the following condition holds:
\begin{align}\label{eq:approximate-polynomial-verification}
        \max_{x\in[-1,1]} \left|\beta p(x)-e^{-\i \alpha x t}\right| \leq \epsilon.
    \end{align}
    Of course, our verification cannot enumerate every element in $[-1,1]$, but since the exponential function can be well-approximated by a polynomial, we can check numerically whether they satisfy (\ref{eq:approximate-polynomial-verification}) on the Chebyshev nodes $\{x_m=\cos\left(\frac{2m-1}{2M}\pi\right):m\in[M]\}$ for $M=O(|\alpha t|+\ln(1/\epsilon)+d)$ and $[M]=\{1,\ldots,M\}$. 
The verification then checks at every grid point $x_m$, $|\beta^*p(x_m)-e^{-\i \alpha x_m t}|\leq \epsilon/2$.
    This requires at most $O(M)$ evaluations of the functions, and can be done efficiently with a classical computer. 
    Although the pointwise threshold $\epsilon/2$ is stricter than the desired uniform error bound, Theorem~\ref{thm:lipschitz-chebyshev} shows that, with sufficiently many Chebyshev nodes, the conditions
    $\left|\beta^*p(x_m)-e^{-\i\alpha x_m t}\right|
        \leq \epsilon/2
        \text{ for all }m\in[M]$
    suffice to guarantee \eqref{eq:approximate-polynomial-verification}.

\section{Symbolic Verification of Block Encoding Circuits}\label{sec:block-encoding}

Let the target Hamiltonian be given as a weighted sum of Pauli strings
$H=\sum_{\ell=0}^{M-1} h_\ell P_\ell$,
where $h_\ell\in \mathbb{C}$ and $P_\ell\in \{I,X,Y,Z\}^{\otimes n}$.
We assume that the block-encoding circuit $U$ acts on $r$ ancilla qubits and $n$ system qubits.
Our symbolic verifier checks whether the \emph{all-zero ancilla block}\footnote{For ancilla bit strings $b,b'\in\{0,1\}^r$, the $(b,b')$-block of $U$ is the system operator
$U_{b,b'} := (\bra{b}_{\mathsf{anc}}\otimes I_{\mathsf{sys}})U(\ket{b'}_{\mathsf{anc}}\otimes I_{\mathsf{sys}})$.
Thus, the all-zero ancilla block is $U_{0^r,0^r}$.} of $U$ is proportional to the target Hamiltonian:
\[
(\bra{0^r}_{\mathsf{anc}}\otimes I_{\mathsf{sys}})U(\ket{0^r}_{\mathsf{anc}}\otimes I_{\mathsf{sys}})= H/\alpha
\]
for some subnormalization factor $\alpha$.

Our verification algorithm symbolically executes the input circuit gate by gate.
Instead of tracking amplitudes over the full Hilbert space, it tracks the system operation induced on each ancilla branch.
Concretely, after every circuit prefix, the verifier maintains one symbolic system operator for each ancilla bit string.
At the end of the symbolic execution, the operator associated with the all-zero ancilla branch is compared with the target Hamiltonian, up to an unknown subnormalization factor.

\subsection{Symbolic states}

Our symbolic verifier maintains a branch-indexed symbolic state
\[
    \sigma=\{B_b\}_{b\in\{0,1\}^r},
    \qquad
    B_b=\sum_{P\in\mathcal P_n} c_{b,P}P,\quad
    c_{b,P}\in\mathbb C,\quad
    \mathcal P_n=\{I,X,Y,Z\}^{\otimes n}.
\]
Here $b$ is an ancilla bit string and $B_b$ is a symbolic system operator. The
state represents
\[
    \ket{\Phi}=\sum_{b\in\{0,1\}^r}
    \ket{b}_{\mathsf{anc}}B_b\ket{\psi}_{\mathsf{sys}},
    \qquad
    B_{0^r}=I,\quad B_b=0\text{ for }b\neq0^r\text{ initially}.
\]

\subsection{Gate updates}

We now describe how quantum gates update symbolic states.
The update rules depend only on how a gate interacts with the ancilla register and the system register.
For the Pauli-LCU block-encoding circuits considered here, it suffices to handle gates of the following forms:
\begin{inparaenum}
    \item Ancilla-only gates,
    \item System-only gates, and
    \item Ancilla-controlled system gates.
\end{inparaenum}
These forms cover the standard LCU structure: PREPARE and PREPARE$^\dagger$ act only on the ancilla register, while SELECT consists of ancilla-controlled Pauli operations on the system register.
We therefore do not describe system-controlled ancilla gates or other gate patterns. Their occurrence in this setting is likely either redundant or indicative of a bug in the circuit construction. When such a gate is encountered, the verifier reports it directly instead of attempting to absorb it into the Pauli-LCU abstraction.

\paragraph{Ancilla-only gates.}
An ancilla-only gate only changes how symbolic branches are combined. Suppose
it acts on $t$ ancilla qubits and is represented by
$M\in\mathbb C^{2^t\times 2^t}$. For every fixed assignment $w$ of the remaining
$r-t$ ancilla bits and affected-bit values $u,v\in\{0,1\}^t$, the verifier
updates
\[
    B_{w,u}'=\sum_{v\in\{0,1\}^t} M_{u,v}B_{w,v}.
\]
For $t=1$, this gives
$B_{w,0}'=M_{00}B_{w,0}+M_{01}B_{w,1}$ and
$B_{w,1}'=M_{10}B_{w,0}+M_{11}B_{w,1}$.
In particular,
\[
\begin{aligned}
\texttt{x anc[j]}:\quad
B_{w,0}' &= B_{w,1},
&
B_{w,1}' &= B_{w,0};\\
\texttt{h anc[j]}:\quad
B_{w,0}' &= \frac{B_{w,0}+B_{w,1}}{\sqrt{2}},
&
B_{w,1}' &= \frac{B_{w,0}-B_{w,1}}{\sqrt{2}}.
\end{aligned}
\]
\paragraph{System-only gates.}
A system-only gate acts only on the system register and leaves the ancilla
branch unchanged. Suppose $G=\sum_{Q\in\mathcal P_n} g_QQ$ is represented as a
Pauli polynomial. Then $B_b'=GB_b$ for every $b\in\{0,1\}^r$. If
$B_b=\sum_{P\in\mathcal P_n}c_{b,P}P$, then
$B_b'=\sum_{Q,P\in\mathcal P_n}g_Qc_{b,P}QP$. In the special case where
$G\in\mathcal P_n$ is a Pauli string, each product satisfies
$GP=\omega_{G,P}P'$ for some $\omega_{G,P}\in\{\pm1,\pm \i\}$ and
$P'\in\mathcal P_n$. Hence a system-only Pauli gate only changes Pauli labels
and coefficients, and does not increase the number of Pauli terms. For example,
$\texttt{x sys[i]}$ updates $B_b'=X_iB_b$ for every $b\in\{0,1\}^r$.
\paragraph{Ancilla-controlled system gates.}
An ancilla-controlled system gate applies a system operation depending on the
ancilla control bits. Suppose a set $J$ of ancilla qubits controls a system gate
$G=\sum_{Q\in\mathcal P_n}g_QQ$, and let $t=|J|$. For every fixed assignment
$w$ of the remaining $r-t$ ancilla bits and every control value
$u\in\{0,1\}^t$, the verifier updates
\[
B_{w,u}'=
\begin{cases}
GB_{w,u} & \text{if }u=1^t,\\
B_{w,u} & \text{otherwise.}
\end{cases}
\]
If $B_{w,1^t}=\sum_{P\in\mathcal P_n}c_{w,1^t,P}P$, then
$B_{w,1^t}'=\sum_{Q,P\in\mathcal P_n}g_Qc_{w,1^t,P}QP$, with the phase from
Pauli multiplication absorbed into the coefficient. For $t=1$, the common
controlled-Pauli updates are
\[
\begin{aligned}
\texttt{cx anc[j], sys[i]}:\quad
B_{w,0}' &= B_{w,0},
&
B_{w,1}' &= X_iB_{w,1};\\
\texttt{cz anc[j], sys[i]}:\quad
B_{w,0}' &= B_{w,0},
&
B_{w,1}' &= Z_iB_{w,1}.
\end{aligned}
\]
When $G$ is a Pauli string, the update does not increase the number of Pauli
terms.
\subsection{Final Proportionality Check}\label{sec:proportionality}

After symbolically executing the whole circuit, the verifier obtains a family
of symbolic operators
\(
\{B_b\}_{b\in\{0,1\}^r}.
\)
By construction, the all-zero branch satisfies
\[
B_{0^r}
=
(\bra{0^r}_{\mathsf{anc}}\otimes I_{\mathsf{sys}})
U
(\ket{0^r}_{\mathsf{anc}}\otimes I_{\mathsf{sys}}).
\]
Therefore, the verifier accepts if the all-zero branch is equal to the target
Hamiltonian up to a subnormalization factor:
\(
B_{0^r}=\frac{H}{\alpha}
\)
for some nonzero scalar $\alpha$.
Equivalently, write
\begin{align}\label{eq:pauli-expansion-hb}
H=\sum_{P\in\mathcal P_n} h_P P
\quad\text{and}\quad
B_{0^r}=\sum_{P\in\mathcal P_n} c_P P.
\end{align}
Since the Pauli strings in $\mathcal P_n$ form a basis for the vector space of
linear operators on $n$ qubits, these Pauli expansions are unique. Hence
checking
\(
B_{0^r}=H/\alpha
\)
is equivalent to checking whether the coefficient vector $c=(c_P)_P$ is
proportional to the coefficient vector $h=(h_P)_P$:
\[
\alpha c_P=h_P
\qquad
\text{for every }P\in\mathcal P_n.
\]

In the implementation of \textsc{Proportional}$(B,H,\varepsilon)$, coefficients are represented numerically. The verifier therefore checks approximate proportionality by finding a positive real scalar $\alpha$ such that $\alpha B_{0^r}$ is close to $H$. More precisely, $\alpha B_{0^r}-H=\sum_{P\in\mathcal P_n}(\alpha c_P-h_P)P$. Since the Pauli strings are orthogonal under the Hilbert--Schmidt inner product and satisfy $\operatorname{tr}(P^\dagger P)=2^n$, we have $\|\alpha B_{0^r}-H\|_F=2^{n/2}\|\alpha c-h\|$. Hence $\|\alpha B_{0^r}-H\|\le \|\alpha B_{0^r}-H\|_F=2^{n/2}\|\alpha c-h\|$. It therefore suffices to check $\|\alpha c-h\|\le \frac{\varepsilon}{2^{n/2}}.$
The verifier infers the subnormalization factor by solving $\min_{\alpha\in\mathbb R_{>0}}\|\alpha c-h\|$. 
To this end, the following lemma is useful.
\begin{lemma}\label{lem:minimization}
    For integer $N\geq 1$, let $a,b$ be two vectors of $N$ entries,
    with $a\neq 0$, and let $C(\gamma)=\|\gamma a - b\|$.
    Then $C$ is minimized at $\gamma=\langle a,b\rangle/\|a\|^2$.
    If $\gamma$ is restricted to be in the set of real numbers, then $C$ is minimized at $\gamma=\mathrm{Re}\langle a,b\rangle/\|a\|^2$. 
\end{lemma}
\begin{proof}
The lemma can be proved by a geometric argument.
Let $b=c_0 a + c_1 a^\bot$ for some vector $a^\bot$ orthogonal to $a$.
This decomposition is always possible since we can choose $a^\bot$ in the span of $\{a,b\}$. 
By taking an inner product with $a$ on both sides, we have the coefficient $c_0=\langle a,b\rangle/\|a\|^2$.
Since the two components are orthogonal,
\[
    \|\gamma a-b\|^2
    =
    |\gamma-c_0|^2\|a\|^2
    +
    \|c_1a^\bot\|^2    
    =
    |\gamma-c_0|^2\|a\|^2
    +
    \|b-c_0a\|^2.
\]
The second term is independent of $\gamma$. Hence the minimum over
$\gamma\in\mathbb C$ is attained at $\gamma=c_0$, while the minimum over
$\gamma\in\mathbb R$ is attained at $\gamma=\operatorname{Re}(c_0)$.

\end{proof}

In both normalization checks, we return $(\Reject,\bot)$ if the
denominator cannot be confirmed to be positive before division.
Geometrically, this projects $h$ onto the ray spanned by $c$. 
If the lower bound of $\|c\|$ is not positive, \textsc{Proportional} returns $(\Reject,\bot)$. 
By \lem{minimization}, the best real subnormalization factor is 
\begin{align}\label{eq:alpha}
\alpha_*=\frac{\mathrm{Re}(\langle c,h\rangle)}{\|c\|^2}.    
\end{align}
If $\alpha_*\le 0$, \textsc{Proportional} returns $(\Reject,\bot)$. Otherwise, it accepts when 
\begin{align}\label{eq:proportionality-check}
    \|\alpha_*c-h\|\le \frac{\varepsilon}{2^{n/2}}.
\end{align}
where $\operatorname{Re}$ takes the real part of a complex number.
\textsc{Proportional}$(B,H,\varepsilon)$ returns a pair $(\Accept,\alpha_*)$ iff $\alpha_* > 0$ and (\ref{eq:proportionality-check}) holds. Otherwise, it returns $(\Reject,\bot)$.

\subsection{Algorithm}
We summarize the symbolic execution procedure above as Algorithm~\ref{alg:verify-block-encoding}.
The algorithm initializes one symbolic branch for each ancilla bit string, processes the input circuit gate by gate using the symbolic update rules, and finally checks whether the all-zero branch is proportional to the target Hamiltonian.
Because the subnormalization factor is not fixed a priori, the final check infers the factor $\alpha$ from the Pauli coefficients.
\begin{algorithm}[ht]
\DontPrintSemicolon
\caption{$\VerifyBlockEncoding(H,C,\epsilon)$}
\label{alg:verify-block-encoding}

\KwIn{
Target Hamiltonian $H=\sum_{P\in\mathcal P_n} h_P P$;
QASM circuit $C$ acting on $r$ ancilla qubits and $n$ system qubits;
tolerance $\varepsilon$
}
\KwOut{
$(\Accept,\alpha)$ if $B_{0^r}$ is a Hermitian satisfying $\|\alpha B_{0^r}-H\|\le\varepsilon$; otherwise $(\Reject,\bot)$.

}

Initialize $B_b\gets 0$ for all $b\in\{0,1\}^r$, and set $B_{0^r}\gets I$\;

\ForEach{instruction $g$ in $C$}{
    Update $\{B_b\}_{b\in\{0,1\}^r}$ according to the symbolic update rule for $g$\;
}

Write
$
H=\sum_{P\in\mathcal P_n} h_P P,
\qquad
B_{0^r}=\sum_{P\in\mathcal P_n} c_P P.
$\;
\lIf{$B_{0^r}\neq B_{0^r}^\dagger$}{Return $(\Reject,\bot)$}
\Return \textsc{Proportional}$(B_{0^r},H,\varepsilon)$\; 

\end{algorithm}
Here $\textsc{Proportional}(B,H,\varepsilon)$ denotes the numerical
proportionality check on Pauli expansions. 
First, write the two matrices as a linear combination of Pauli matrices, as in (\ref{eq:pauli-expansion-hb}).
The check treats $h=(h_P)_P$ and $c=(c_P)_P$ as coefficient vectors and infers
the positive real subnormalization factor $\alpha$ that best fits
\(
h\approx \alpha c.
\)
Equivalently, as analyzed in Section~\ref{sec:proportionality}, 
it computes the inferred value $\alpha_*$ as in (\ref{eq:alpha}).
The verifier rejects if
$\alpha_*\le 0$ or the principal block is not Hermitian, and otherwise checks if (\ref{eq:proportionality-check}) holds.
When this check succeeds, the verifier accepts $B$ as a block encoding of $H$
with the subnormalization factor $\alpha_*$.

\paragraph{\bfseries Branch complexity.}
If the block-encoding circuit uses $r$ ancilla qubits, symbolic execution
tracks one Pauli-polynomial system operator $B_b$ for each branch
$b\in\{0,1\}^r$, hence $2^r$ symbolic branches. For a standard Pauli-LCU block
encoding of $M$ Pauli terms, $r=\lceil\log_2 M\rceil$, so the number of
branches is $O(M)$.
The size of each branch expression is circuit-dependent. In the worst case,
symbolic multiplication can make a branch expression grow exponentially, and
we do not claim a polynomial worst-case bound for arbitrary circuits. However,
the verifier avoids constructing the exponentially large system matrix: it
tracks Pauli-polynomial expressions attached to ancilla branches. In our
evaluation, this representation is sufficient to verify block encodings of
structured Pauli-sum Hamiltonians, including Ising, MaxCut, and Heisenberg
instances, with hundreds of Pauli terms and up to 256 system qubits.

\begin{theorem}[Soundness of the block-encoding verifier]
\label{thm:soundness-verify-block-encoding}
Let $U$ be the unitary implemented by the input circuit $C$.
Assume that $C$ uses only the gate forms supported by the symbolic semantics
in Section~\ref{sec:block-encoding}.
If Algorithm~\ref{alg:verify-block-encoding} returns $(\Accept,\alpha)$
on input $(H,C,\varepsilon)$, then $\alpha>0$,the
principal block
$
    B_{0^r}
    =
    (\langle 0^r|_{\mathrm{anc}}\otimes I_{\mathrm{sys}})
    U
    (|0^r\rangle_{\mathrm{anc}}\otimes I_{\mathrm{sys}})
$ is Hermitian, and
$\|\alpha B_{0^r}-H\|\leq\varepsilon.$

\end{theorem}

\begin{proof}
The symbolic execution rules maintain the invariant that after any circuit prefix $C_t$ with denotation $U_t$, the symbolic branches satisfy $U_t(\ket{0^r}_{\mathsf{anc}}\ket{\psi}_{\mathsf{sys}})=\sum_{b\in\{0,1\}^r}\ket{b}_{\mathsf{anc}}B_b^{(t)}\ket{\psi}_{\mathsf{sys}}$ for every system state $\ket{\psi}$. This holds initially by $B_{0^r}^{(0)}=I$ and $B_b^{(0)}=0$ for $b\neq 0^r$, and each update rule is obtained by applying the corresponding gate to this branch decomposition. Hence, after processing the whole circuit, $B_{0^r}=(\bra{0^r}_{\mathsf{anc}}\otimes I_{\mathsf{sys}})U(\ket{0^r}_{\mathsf{anc}}\otimes I_{\mathsf{sys}})$.

It remains to use the final proportionality check. Write $H=\sum_{P\in\mathcal P_n}h_PP$ and $B_{0^r}=\sum_{P\in\mathcal P_n}c_PP$, and let $h=(h_P)_P$ and $c=(c_P)_P$. When the algorithm accepts, \textsc{Proportional} returns a positive real $\alpha$ such that $\|\alpha c-h\|\le \varepsilon/2^{n/2}$. Let $D=\alpha B_{0^r}-H=\sum_{P\in\mathcal P_n}(\alpha c_P-h_P)P$. By orthogonality of the $n$-qubit Pauli basis, $\|D\|_F=2^{n/2}\|\alpha c-h\|$. Since $\|D\|\le \|D\|_F$, we get $\|\alpha B_{0^r}-H\|\le 2^{n/2}\|\alpha c-h\|\le \varepsilon$. Substituting the expression for $B_{0^r}$ proves the claim. Moreover, Algorithm~1 accepts only if $B_{0^r}=B_{0^r}^\dagger$, which establishes the Hermiticity claim.
\end{proof}

\section{Symbolic Execution of Quantum Signal Processing}
\label{sec:symbolic-qsp}

This section describes the symbolic execution of quantum signal processing
(QSP) circuits built from a certified block encoding. A QSP circuit repeatedly
invokes a block-encoding unitary $U_H$ and its adjoint $U_H^\dagger$.
Rather than expanding the gate-level implementation of $U_H$ at every
invocation, the verifier reuses the block-encoding certificate as an abstract
interface and symbolically executes the surrounding QSP circuit.
Throughout this section, we follow the notation of
Section~\ref{sec:block-encoding}: the system register contains
$n$ qubits, and the certified block encoding uses $r$ block-encoding ancilla
qubits. In addition, the QSP layer uses $s$ phase ancilla qubits. Thus the
circuit has three disjoint registers: $s$ phase-ancilla qubits, $r$
block-encoding ancilla qubits, and $n$ system qubits. We index phase-ancilla
branches by $p\in\{0,1\}^s$, and block-encoding ancilla branches by
$b\in\{0,1\}^r$.

The purpose of this phase is to compute the symbolic expression implemented by
the principal QSP branch. The verifier tracks the phase-ancilla branches
explicitly. For the internal block-encoding ancilla, however, it distinguishes
only the all-zero branch $0^r$ from the aggregate of all nonzero branches
$b\neq 0^r$. Each invocation of $U_H$ or $U_H^\dagger$ updates the current
symbolic expressions according to this two-component abstraction. After all QSP
gates have been processed, the principal component of the all-zero phase branch
contains the expression produced by symbolic execution. This expression is then
passed to the polynomial-extraction phase.

\subsection{Four-Block Matrix Abstraction of Block Encodings}
\label{subsec:four-block-abstraction}

Once a block-encoding circuit $U_H$ has been certified, the verifier no longer
expands its internal gate structure. Instead, it replaces $U_H$ by a typed
four-block matrix abstraction. Since the system register has $n$ qubits, the
principal all-zero block-encoding branch has dimension $2^n$. The aggregate of
all nonzero block-encoding branches has dimension $(2^r-1)2^n$.

With respect to the decomposition into the all-zero branch and the aggregate
nonzero branch, we write
\[
    U_H =
    \begin{bmatrix}
        H & G\\
        A & C
    \end{bmatrix},
    \qquad
    U_H^\dagger =
    \begin{bmatrix}
        H^\dagger & A^\dagger\\
        G^\dagger & C^\dagger
    \end{bmatrix}.
\]
Here the four matrix variables have dimensions
\[
\begin{aligned}
    H &\in \mathbb C^{2^n\times 2^n},
    &
    G &\in \mathbb C^{2^n\times (2^r-1)2^n},&
    A &\in \mathbb C^{(2^r-1)2^n\times 2^n},
    &
    C &\in \mathbb C^{(2^r-1)2^n\times (2^r-1)2^n}.
\end{aligned}
\]
The principal block $H$ is the block certified by the block-encoding verifier:
\[
    H=(\langle 0^r|\otimes I)U_H(|0^r\rangle\otimes I).
\]
In the first verification phase, this block is proved to equal the target
Hamiltonian up to a subnormalization factor. In the Hamiltonian-simulation
phase, this subnormalization factor is absorbed into the effective simulation
time. Therefore, in the QSP symbolic execution below, we use $H$ to denote the
normalized encoded Hamiltonian block.

The remaining blocks $A$, $G$, and $C$ summarize the behavior of $U_H$
involving the aggregate nonzero block-encoding branches. They are treated as
formal matrix variables with the dimensions above and are not expanded into the
internal gates of the certified block encoding.

This abstraction is used only after the block encoding has been verified. It
therefore separates the verification of the QSP layer from the gate-level
verification of the block-encoding circuit: the QSP verifier sees only the
certified principal block $H$, the formal garbage blocks $A,G,C$, and the fact
that these blocks arise from a unitary matrix.

\subsection{Symbolic Execution over the Four-Block Abstraction}
\label{subsec:qsp-symbolic-execution}

We now describe how the verifier symbolically executes a QSP circuit after
replacing each call to $U_H$ by its four-block matrix abstraction.
The symbolic executor treats the three registers differently. The phase ancilla
register is tracked explicitly: each branch $p\in\{0,1\}^s$ carries its own
symbolic pair. The block-encoding ancilla register is tracked only through the
two-component abstraction introduced above: the all-zero branch and the
aggregate of all nonzero branches. Finally, the system register is represented
symbolically through the matrix expressions attached to these components.

The four-block abstraction is valid only for QSP circuits that preserve this
two-component view of the block-encoding ancilla. Between two consecutive
invocations of the certified block encoding, no gate may modify the system
register directly. Moreover, operations on the block-encoding ancilla are
restricted to those whose behavior depends only on whether the ancilla is in
the all-zero state. For example, multi-controlled gates conditioned on the
all-zero block-encoding ancilla state are supported, whereas operations that
distinguish individual nonzero block-encoding ancilla states are not.
Under these restrictions, the verifier never needs to distinguish individual
nonzero block-encoding ancilla branches during QSP symbolic execution. For
each phase-ancilla branch, it maintains only two symbolic components: the
principal component corresponding to the all-zero block-encoding branch, and
the aggregate non-principal component representing all remaining
block-encoding branches.

\subsection{Symbolic States}

For each phase-ancilla branch $p\in\{0,1\}^s$, the verifier maintains a pair
$(B_{p,0},B_{p,\neq0})$. The expression $B_{p,0}$ represents the component in
the all-zero block-encoding branch, while $B_{p,\neq0}$ represents the aggregate
component over all nonzero block-encoding branches. Their dimensions are
\[
    B_{p,0}\in\mathbb C^{2^n\times 2^n},
    \qquad
    B_{p,\neq0}\in\mathbb C^{(2^r-1)2^n\times 2^n}.
\]
Thus $B_{p,0}$ and $B_{p,\neq0}$ have the same input dimension but different
output dimensions. The symbolic state is
$\mathcal B_{\mathrm{QSP}}=\{(B_{p,0},B_{p,\neq0})\}_{p\in\{0,1\}^s}$, initialized by
\[
    B_{0^s,0}=I,
    \qquad
    B_{p,0}=0 \text{ for } p\neq 0^s,
    \qquad
    B_{p,\neq0}=0 \text{ for all } p\in\{0,1\}^s.
\]
The verifier then symbolically executes the QSP circuit gate by gate.
\subsection{Aggregate-Compatible Gate Updates}
\label{subsec:qsp-aggregate-compatible-updates}

We now describe the gate updates used by the QSP symbolic executor. The updates
are restricted to gates that preserve the aggregate abstraction introduced
above. The verifier may explicitly distinguish computational-basis branches of
the phase ancilla register, but it may not distinguish individual nonzero
branches of the block-encoding ancilla register.

\paragraph{Phase-ancilla gates.}
Gates acting only on the phase ancilla register are handled branch-wise. Since
the phase branches are tracked explicitly, an ancilla-only gate on the phase
register mixes the symbolic pairs
$(B_{p,0},B_{p,\neq0})$ across different branches $p\in\{0,1\}^s$ according to
the usual matrix action of the gate. The same branch mixing is applied
simultaneously to the principal components and to the aggregate non-principal
components.

For example, under the convention
$
R_z(\theta)=e^{-i\theta Z/2}
=
\begin{bmatrix}
e^{-i\theta/2} & 0\\
0 & e^{i\theta/2}
\end{bmatrix}$,
applying $R_z(\theta)$ to a phase-ancilla qubit multiplies the symbolic pair in
each branch with that qubit equal to $0$ by $e^{-i\theta/2}$, and the symbolic
pair in each branch with that qubit equal to $1$ by $e^{i\theta/2}$.

\paragraph{Operations controlled by the phase ancilla.}
Some QSP circuits contain operations controlled by one or more phase-ancilla
qubits. Since phase-ancilla branches are tracked explicitly, such operations
are handled branch by branch. For a fixed phase branch $p$, if the control
condition is not satisfied, the pair $(B_{p,0},B_{p,\neq0})$ is left unchanged.
If the control condition is satisfied, the corresponding operation is applied
to that pair.
In particular, a controlled call to $U_H$ or $U_H^\dagger$ applies the
four-block update only on the phase branches where the controls are enabled.
The explicit update rules for calls to $U_H$ and $U_H^\dagger$ are given in
Section~\ref{subsubsec:qsp-block-encoding-calls}.

\paragraph{Operations on the block-encoding ancilla.}
Operations that access the internal block-encoding ancilla are allowed only if
they preserve the distinction between the all-zero branch and the aggregate
nonzero component. Such operations may test whether the block-encoding ancilla
is in the all-zero state, but they may not distinguish among different nonzero
block-encoding ancilla states.
For example, a phase operation conditioned on the block-encoding ancilla being
in the all-zero state is represented by
$B'_{p,0}=\omega B_{p,0}$ and $B'_{p,\neq0}=B_{p,\neq0}$, where $\omega\in\mathbb C$ is the corresponding phase factor. Similarly, an operation that applies the same scalar factor to every nonzero block-encoding
ancilla branch is represented by
$B'_{p,0}=B_{p,0}$ and $B'_{p,\neq0}=\omega B_{p,\neq0}$.
Both updates are well typed because scalar multiplication preserves the
dimensions of blocks.

Operations that distinguish two different nonzero block-encoding ancilla
branches are not supported by this abstraction. Supporting such gates would
require returning to the finer branch representation used by the
block-encoding verifier, where each ancilla branch is tracked
separately.

\paragraph{System-side operations.}
Between two consecutive invocations of the certified block encoding, the QSP
gate language does not allow gates that modify the system register directly.
All system-side transformations in the QSP layer are introduced through calls
to $U_H$ and $U_H^\dagger$, which are handled by the four-block abstraction in
Section~\ref{subsubsec:qsp-block-encoding-calls}. This restriction ensures
that the aggregate non-principal component can be treated as a single symbolic
component without expanding the individual nonzero block-encoding ancilla
branches.

\subsection{Calls to the Block Encoding}
\label{subsubsec:qsp-block-encoding-calls}

A call to the certified block encoding $U_H$ or its adjoint $U_H^\dagger$ acts
on the block-encoding ancilla and the system register. In the QSP verifier,
this call is not expanded into the gate-level implementation of $U_H$; instead,
it is executed using the four-block abstraction.

For each fixed phase-ancilla branch $p\in\{0,1\}^s$, the symbolic state contains
a pair $(B_{p,0},B_{p,\neq0})$. A call to $U_H$ updates this pair by
$\begin{bmatrix}
B'_{p,0}\\
B'_{p,\neq0}
\end{bmatrix}
=
\begin{bmatrix}
H & G\\
A & C
\end{bmatrix}
\begin{bmatrix}
B_{p,0}\\
B_{p,\neq0}
\end{bmatrix}$,
or equivalently,
$B'_{p,0}=HB_{p,0}+GB_{p,\neq0}$ and
$B'_{p,\neq0}=AB_{p,0}+CB_{p,\neq0}$. These products are well typed: both
$HB_{p,0}$ and $GB_{p,\neq0}$ are $2^n\times 2^n$ matrices, while both
$AB_{p,0}$ and $CB_{p,\neq0}$ are $(2^r-1)2^n\times 2^n$ matrices.

A call to $U_H^\dagger$ is handled similarly:
$\begin{bmatrix}
B'_{p,0}\\
B'_{p,\neq0}
\end{bmatrix}
=
\begin{bmatrix}
H^\dagger & A^\dagger\\
G^\dagger & C^\dagger
\end{bmatrix}
\begin{bmatrix}
B_{p,0}\\
B_{p,\neq0}
\end{bmatrix}$.
Equivalently,
$B'_{p,0}=H^\dagger B_{p,0}+A^\dagger B_{p,\neq0}$ and
$B'_{p,\neq0}=G^\dagger B_{p,0}+C^\dagger B_{p,\neq0}$. Again, both terms in
the first sum are $2^n\times 2^n$ matrices, and both terms in the second sum
are $(2^r-1)2^n\times 2^n$ matrices.

If the call is controlled by one or more phase-ancilla qubits, then the update
is applied only to the phase branches on which the controls are enabled. For
all other phase branches, the pair $(B_{p,0},B_{p,\neq0})$ is left unchanged.
For example, if a call to $U_H$ is controlled on a phase-ancilla qubit being
$1$, then every branch $p$ with that qubit equal to $1$ receives the $U_H$
update above, while all branches $p$ with that qubit equal to $0$ are unchanged.

\begin{algorithm}[t]
\DontPrintSemicolon
\caption{$\SymbolicExecute(C)$}
\label{alg:symbolic-execute}
\KwIn{QSP circuit $C$, number of phase ancilla $s$}
\KwOut{Symbolic expression $E$}

Initialize $B_{0^s,0}\gets I$ and
$B_{p,\eta}\gets 0$ for
$(p,\eta)\in\{0,1\}^s\times\{0,\neq0\}\setminus\{(0^s,0)\}$\;

\ForEach{instruction $g$ in $C$}{
    Update
    $\{B_{p,\eta}\}_{p\in\{0,1\}^s,\ \eta\in\{0,\neq0\}}$
    according to the symbolic update rule for $g$\;

    Normalize each $B_{p,\eta}$ with the rewrite rules of \sec{qsp-polynomial-extraction}.
}

\Return $B_{0^s,0}$
\end{algorithm}

\begin{theorem}[Correctness of symbolic execution]
\label{thm:qsp-symbolic-execution}
Let $C$ be a QSP circuit supported by the symbolic semantics, and let $U_C$
be its unitary. If $E=\textsc{SymbolicExecute}(C)$, then
$(\llbracket E\rrbracket
=
(\bra{0^s}\otimes\bra{0^r}\otimes I)
U_C
(\ket{0^s}\otimes\ket{0^r}\otimes I))$.
Hence, Algorithm~\ref{alg:symbolic-execute} returns exactly the principal block of the QSP circuit.
\end{theorem}

\begin{proof}
We induct on the number of executed instructions.
For each phase branch $p$, the interpreted pair
$(B_{p,0},B_{p,\neq0})$ equals the principal and non-principal
components of the circuit prefix applied to
$\ket{0^s}\otimes\ket{0^r}\otimes I$.
The initialization satisfies this property.
Phase-ancilla gates mix these components according to their
matrix entries. Phase operations conditioned on the zero or
nonzero block-encoding branch multiply the corresponding
components by the specified scalars.
Calls to $U_H$ and $U_H^\dagger$ apply their exact four-block
matrix action. Phase-controlled operations apply the relevant
update only on enabled phase branches.
Thus every supported instruction preserves the property.
After the last instruction, $B_{0^s,0}$ is exactly the
principal block stated in the theorem.
\end{proof}

\section{Polynomial Extraction by Unitarity Simplification}
\label{sec:qsp-polynomial-extraction}

After symbolic execution, the verifier obtains a symbolic state
$\mathcal B_{\mathrm{QSP}}
=\{(B_{p,0},B_{p,\neq0})\}_{p\in\{0,1\}^s}$. The operator implemented by the
QSP circuit is represented by the principal component of the all-zero phase
branch, namely $B_{0^s,0}$.
The goal of this stage is to extract the polynomial transformation implemented
by the QSP circuit. The raw expression $B_{0^s,0}$ may contain the auxiliary
blocks $A,G,C$ and their adjoints, corresponding to intermediate transitions
through the aggregate non-principal component. The verifier applies a fixed
oriented rewrite system derived from the unitarity of the four-block
abstraction. If the normalized expression contains only $H$, scalar
coefficients, and identities, the verifier reads it as a polynomial in $H$.
Otherwise, the verifier rejects.

\paragraph{Dimension compatibility.}
The block variables $H,G,A,C$ have the matrix dimensions specified in
Section~\ref{subsec:four-block-abstraction}. Symbolic execution generates
expressions by dimension-compatible block-matrix multiplication through calls
to $U_H$ and $U_H^\dagger$: products such as $GA$, $HG$, and $AH$ may arise,
whereas dimension-incompatible products such as $HA$ are not generated. The polynomial extractor applies rewrite rules only to dimension-compatible
subexpressions.

\paragraph{Rewrite rules from unitarity.}
Recall that $U_H$ is unitary. Hence the block identities induced by $U_H^\dagger U_H=I$ and
$U_HU_H^\dagger=I$ are valid. The verifier orients these identities into rewrite rules:
\[
\begin{array}{llll}
\mathrm{R1}:& A^\dagger A \to I-H^\dagger H,
&
\mathrm{R2}:& GG^\dagger \to I-HH^\dagger,\\
\mathrm{R3}:& A^\dagger C \to -H^\dagger G,
&
\mathrm{R4}:& GC^\dagger \to -HA^\dagger,\\
\mathrm{R5}:& C^\dagger A \to -G^\dagger H,
&
\mathrm{R6}:& CG^\dagger \to -AH^\dagger,\\
\mathrm{R7}:& G^\dagger G \to I-C^\dagger C,
&
\mathrm{R8}:& AA^\dagger \to I-CC^\dagger .
\end{array}
\]
Each rule is a dimension-compatible matrix identity. For example,
$A^\dagger A$ and $I-H^\dagger H$ have the same dimensions, as do
$A^\dagger C$ and $-H^\dagger G$.
Rules R1--R6 reduce the number of non-principal block occurrences. Rules R7
and R8 preserve the total number of non-principal block occurrences, but reduce
the number of occurrences of $A$ and $G$. The rewrite system is sound but
intentionally incomplete: if the deterministic simplifier does not reduce the
expression to a polynomial in $H$, the verifier rejects.

\paragraph{Algebraic normalization.}
In addition to the unitarity rules, the verifier applies the following
algebraic normalization rules:
\[
\begin{array}{llll}
\mathrm{D1}:& X(\beta Y+\kappa Z)\to \beta XY+\kappa XZ,
&
\mathrm{D2}:& (\beta Y+\kappa Z)X\to \beta YX+\kappa ZX,\\
\mathrm{S1}:& IX\to X,\quad XI\to X,
&
\mathrm{S2}:& 0X\to0,\quad X0\to0,\quad X+0\to X,\\
\mathrm{S3}:& \beta X+\kappa X\to(\beta+\kappa)X,
&
\mathrm{S4}:& HH^k\to H^{k+1},\quad H^kH\to H^{k+1},\\
\mathrm{H}:& H^\dagger\to H.
\end{array}
\]
Here $\beta,\kappa\in\mathbb C$, and rule H is used only in the Hamiltonian
simulation setting, where the certified principal block $H$ is Hermitian. The
implementation also flattens associative sums and products before applying
these rules. These algebraic rules are ordinary matrix-algebra simplifications
and do not use any additional property of the block encoding. Rules H and S1--S4 are applied after flattening associative sums and products, so
they should be read as part of the canonicalization procedure rather than as a
complete rewrite system.

\paragraph{Canonicalization and termination.}
Rules H and S1--S4 are implemented as a deterministic canonicalization pass: the
verifier flattens sums/products, moves scalars to the front, combines identical
monomials, collects powers of $H$, and removes zero terms. The simplifier
applies rules in the fixed order R1--R8, then D1--D2, interleaved with this
canonicalization. To see why this process terminates, imagine fully
expanding the expression into a sum of monomials, without collecting
coefficients. Distribution does not change these expanded terms.
Each R step replaces affected terms by finitely many simpler terms:
each new term has fewer non-principal blocks, or the same number
but fewer occurrences of $A,A^\dagger,G,G^\dagger$.
These counts cannot decrease indefinitely. Since each replacement
produces only finitely many terms, only finitely many R steps are possible.
Between R steps and after the last one, distribution finishes expanding
a finite expression; it never factors sums back into products.
Canonicalization also terminates and only simplifies or combines terms.
Hence the whole process terminates.

\begin{algorithm}[t]
\DontPrintSemicolon
\caption{$\ExtractPolynomial(E)$}
\label{alg:extract-polynomial}
\KwIn{Dimension-compatible symbolic expression $E$}
\KwOut{$(\Accept,p)$ if $E$ rewrites to a polynomial $p(H)$; $(\Reject,\bot)$ otherwise}

\While{some rewrite or algebraic normalization rule applies to $E$}{
    Update $E$ using the first applicable rule in the fixed priority order\;
}

\lIf{$E$ contains any symbol in $\{G,G^\dagger,A,A^\dagger,C,C^\dagger\}$}{
    \Return $(\Reject,\bot)$
}
\lElse{
    collect the coefficients of the powers of $H$ in $E$ and return $(\Accept,p)$
}
\end{algorithm}

\subsection{Example: Symbolic Execution and Polynomial Extraction}

\begin{wraptable}{r}{0.45\textwidth}
\vspace{-5mm}
\setlength{\tabcolsep}{12pt}
\begin{tabular}{@{}cl@{}}
\toprule
rule & expression \\
\midrule
    & $\displaystyle H(-H^\dagger H + A^\dagger A) + G(G^\dagger H - C^\dagger A)$ \\
R1 & $\displaystyle H(-2H^\dagger H + I) + G(G^\dagger H - C^\dagger A)$ \\
R5 & $\displaystyle H(-2H^\dagger H + I) + G(2G^\dagger H)$ \\
D1 & $\displaystyle H(-2H^\dagger H + I) + 2GG^\dagger H$ \\
R2 & $\displaystyle H(-2H^\dagger H + I) + 2(I - HH^\dagger)H$ \\
D1 & $\displaystyle -2HH^\dagger H + H + 2(I - HH^\dagger)H$ \\
D2 & $\displaystyle -2HH^\dagger H + H + (2H - 2HH^\dagger H)$ \\
H   & $\displaystyle -4H^3 + 3H$ \\
\bottomrule
\end{tabular}
\caption{Simplifying the symbolic expression.}
\vspace{-8mm}

\label{tab:example-rewrites}
\end{wraptable}
To better explain how our abstraction works, consider the following example for symbolic execution of a QSP circuit.
For demonstrating the usage of QSP and our verification, we simplify our example so the resulting polynomial is not particularly attached to a Hamiltonian simulation. 

Consider the following implementation of quantum singular value transform based on quantum signal processing:
We invoke our verification scheme to prove that it produces a block encoding of $p(H)=-4H^3+3H$, where $H$ is block-encoded by $U_H$ with subnormalization factor $\alpha=1$.
This effectively reduces each application of $U_H$ or its Hermitian conjugate $(U_H)^\dag$ as a single-qubit abstract gate, illustrated as the following quantum circuit.

\begin{figure}  [tbp]
\begin{center}
    \begin{quantikz}[row sep=0.1cm, column sep=0.15cm]
\lstick{\(q_0\)} 
& \qw & \qw & \targ{} & \gate{R_z(\pi)} & \targ{} & \qw & \qw & \targ{} & \gate{R_z(\pi)} & \targ{} & \qw  & \qw & \qw \\

\lstick{\(q_1\)} 
& \gate[1]{\begin{bmatrix}
    H & G \\ A & C
\end{bmatrix}} & \qw & \octrl{-1} & \qw & \octrl{-1} & \gate{\begin{bmatrix}
    H^\dag & A^\dag \\ G^\dag & C^\dag
\end{bmatrix}} & \qw & \octrl{-1} & \qw & \octrl{-1} & \gate[2]{\begin{bmatrix}
    H & G \\ A & C
\end{bmatrix}} & \qw & \qw \\
\end{quantikz}
\end{center}
\caption{An abstraction of the circuit in Figure~\ref{fig:example-qsp} for symbolic execution.}
\label{fig:example-qsp-abstraction}
\end{figure}

We perform symbolic execution on the circuit in Fig.~\ref{fig:example-qsp-abstraction}:
starting from input block of the form $(I,0)$ representing the state on $q_1$ (with that of $q_0$ initialized to $\ket{0}$) and recalling that $R_z(\theta)=e^{-\i\theta Z/2}$, we proceed the symbolic execution and summarize the steps in Table~\ref{tab:example-symbolic-execution}.
\begin{table}[htb]
\vspace{-3mm}
\begin{center}\small
\scalebox{0.75}{
\begin{tabular}{l|cccc}\toprule
    $op$ & $00$ & $01$ & $10$ & $11$ \\
    \midrule
         & $I$ & $0$ & $0$ & $0$ \\
    $U_H$ & $H$ & $A$ & $0$ & $0$ \\
    $CX$ & $0$ & $A$ & $H$ & $0$ \\
    $R_z$ & $0$ & $-\i A$ & $\i H$ & $0$ \\
    $CX$ & $\i H$ & $-\i A$ & $0$ & $0$ \\
    $(U_H)^\dag$ 
        & $\i H^\dag H - \i A^\dag A$ 
        & $\i G^\dag H - \i C^\dag A$ 
        & $0$ 
        & $0$ \\
    $CX$ 
        & $0$ 
        & $\i G^\dag H - \i C^\dag A$ 
        & $\i H^\dag H - \i A^\dag A$ 
        & $0$ \\
    $R_z$ 
        & $0$ 
        & $G^\dag H - C^\dag A$ 
        & $-H^\dag H + A^\dag A$ 
        & $0$ \\
    $CX$ 
        & $-H^\dag H + A^\dag A$ 
        & $G^\dag H - C^\dag A$ 
        & $0$ 
        & $0$ \\
    $U_H$ 
        & $H(-H^\dag H + A^\dag A) + G(G^\dag H - C^\dag A)$ 
        & $A(-H^\dag H + A^\dag A) + C(G^\dag H - C^\dag A)$ 
        & $0$ 
        & $0$ \\
    \bottomrule
\end{tabular}}
\end{center}
\caption{The symbolic execution of the abstract circuit in Fig.~\ref{fig:example-qsp-abstraction}.}
\vspace{-3mm}
\label{tab:example-symbolic-execution}
\end{table}

\begin{wrapfigure}{R}{0.6\textwidth}
\vspace{-5mm}
\begin{center}
\begin{quantikz}[row sep=0.1cm, column sep=0.15cm]
\lstick{\(q_0\)} 
& \qw & \qw & \targ{} & \gate{R_z(\pi)} & \targ{} & \qw & \qw & \targ{} & \gate{R_z(\pi)} & \targ{} & \qw  & \qw & \qw \\

\lstick{\(q_1\)} 
& \gate[2]{U_H} & \qw & \octrl{-1} & \qw & \octrl{-1} & \gate[2]{U_H^\dagger} & \qw & \octrl{-1} & \qw & \octrl{-1} & \gate[2]{U_H} & \qw & \qw \\

\lstick{\(q_2\)} 
& \qw & \qw & \qw & \qw & \qw & \qw & \qw & \qw & \qw & \qw & \qw & \qw & \qw
    \end{quantikz}\end{center}
\caption{An example QSP circuit.}
\vspace{-3mm}
\label{fig:example-qsp}
\end{wrapfigure}

We will only simplify the expression in the first branch at the end, i.e., the ``$00$'' entry in the last row of Table~\ref{tab:example-symbolic-execution}. 
We then apply rewrite rules to simplify the symbolic expression, as in Table~\ref{tab:example-rewrites}.
Here we omit detailed simplifications of symbolic expressions, and only show crucial steps using the rules from Section~\ref{sec:qsp-polynomial-extraction}.

The above symbolic execution treats the quantum state in each step of the execution as a vector of four symbols, starting from $(I,0,0,0)$.
The $U_H$ is abstracted into a four-block matrix acting on the abstract wire $q_1$, and thus in the joint space of $q_0$ and $q_1$, each application of $U_H$ is an application of the unitary 
\begin{align}
    I\otimes U_H = 
    \begin{bmatrix}
        H & G & 0 & 0 \\
        A & C & 0 & 0 \\
        0 & 0 & H & G \\
        0 & 0 & A & C
    \end{bmatrix}
\end{align}
Similarly, each application of $R_z(\phi)$ acts as a multiplication of the matrix $\mathrm{diag}(e^{-\i\phi/2},e^{-\i\phi/2},e^{\i\phi/2},e^{\i\phi/2})$, which is equivalent to a phase multiplication to each branch. 
The open controlled NOT acts as a swap of the 00 and 10 branches, since it corresponds to a Pauli-$X$ gate when $q_1$ is in the zero state.

Note that in this example, we have shown that the symbolic execution of QSP corresponds to application of one- and two-qubit gates to a symbolic expression of a two-qubit state, and therefore can be simulated efficiently. 
To conclude that the circuit results in a block encoding of a polynomial in $H$, we develop a tool to automatically simplify the symbolic expression in the $00$ branch using the rules that reflect the unitary structure. 

\begin{theorem}\label{thm:soundness-extract-polynomial} The rewrite rules form a sound system. \end{theorem} \begin{proof} We prove the soundness using the fact that $U_H$ is a symbolic expression of a unitary. \[ U_H (U_H)^\dag = \begin{bmatrix} HH^\dag + GG^\dag & HA^\dag + GC^\dag \\ AH^\dag + CG^\dag & AA^\dag + CC^\dag \end{bmatrix} = \begin{bmatrix} I & 0 \\ 0 & I \end{bmatrix} \] This justifies the soundness of R2, R4, R6, and R8. Similarly, since \[ (U_H)^\dag U_H = \begin{bmatrix} H^\dag H + A^\dag A & H^\dag G+A^\dag C \\ G^\dag H + C^\dag A & G^\dag G + C^\dag C \end{bmatrix} = \begin{bmatrix} I & 0 \\ 0 & I \end{bmatrix} \] This justifies the soundness of R1, R3, R5, and R7. The soundness of rule H follows from the fact that $H$ represents a Hermitian matrix. The soundness of D1 and D2 follows from the fact that the symbols represent matrices and from standard matrix arithmetic. \end{proof} \begin{corollary}\label{cor:soundness-extract-polynomial} \alg{extract-polynomial} is sound, i.e., if \alg{extract-polynomial} accepts, then the expression $p$ is proved to be equivalent to the symbolic expression $E$ using the rewrite rules in the logic. \end{corollary} \begin{proof} If \alg{extract-polynomial} accepts, then there must be a polynomial deduced from the rewrite rules. The theorem follows from the soundness of the logic, proved in \thm{soundness-extract-polynomial}. \end{proof}

\subsection{Completeness of Rewrite Rules for Standard QSP Circuits}

In this section, we show that when the quantum program has a standard QSP form, it is always possible to apply our rewrite rules to extract a polynomial in the topleft block $H$. 
First, we give the grammar of a standard QSP circuit.

\begin{definition}[Standard QSP circuits]\label{dfn:well-formed}
    A quantum program is a standard QSP circuit if it has the following patterns:
    \begin{enumerate}
        \item The program alternates its queries to $U$ and $U^\dag$.
        \item Between two adjacent queries to $U$ or $U^\dag$, an arbitrary number of $CX$ gates (as shown in \fig{example-qsp}) can be applied to $q_1q_0$ and that of single-qubit gates can to $q_0$.
    \end{enumerate}
    Formally, the expression $E$ of the quantum program is a standard QSP circuit, if it follows the grammar: 
    \begin{align}
        I &:= \epsilon \mid CX[q_1, q_0]; I \mid W[q_0]; I \\
        A &:= \epsilon \mid U^\dag[q_1, q_2]; I; B \\
        B &:= \epsilon \mid U[q_1, q_2]; I; A\\
        E &:= I \mid I; U[q_1, q_2]; I; A \mid I; U^\dag [q_1, q_2]; I; B.
    \end{align}
    Here $W$ denotes any single-qubit gate.
\end{definition}

The standard QSP circuit defined in \dfn{well-formed} can be easily checked by parsing the program expression. 
Given the definition, we show that when the quantum program is a standard QSP circuit, there is always a composition of the rewrite rules that simplifies the resulting symbolic expression of the abstract state to a polynomial. 
To show this, we define equivalence classes of expressions to make our argument concrete. 
\begin{definition}[Equivalent class of expressions]\label{dfn:equivalent-class}
    We define an equivalence class $\I$ of symbolic expressions as those which can be simplified to a polynomial in $H$ using the rewrite rules.
    We use the notation $M\I$ to denote the class of expressions that are simplified to $Mf(H)$ for some polynomial $f$ for $M\in\{\epsilon,H,G,A,C,H^\dag,G^\dag,A^\dag,C^\dag\}$, where $\epsilon$ is an empty expression. 
    We say that a tuple of symbolic expressions $(v_1,v_2,\ldots)\in (M_1\I, M_2\I,\ldots)$ if each component $v_i\in M_i\I$. 
\end{definition}

\begin{theorem}[Completeness of rewrite rules for standard QSP programs]
    For every standard QSP circuit (as defined in \dfn{well-formed}), in each step of symbolic execution, there exists a composition of rewrite rules that deduces the resulting state to expressions in the class $(\I,M\I,\I,M\I)$ for some $M\in\{A,G^\dag\}$. 
    In particular, the first element of the abstract state is a polynomial in $H$. 
\end{theorem}
\begin{proof}
    Using the convention provided by \dfn{equivalent-class}, we show that after each application of $U[q_1,q_2]$, the resulting state is in $(\I,A\I,\I,A\I)$ and after each application of $U^\dag[q_1,q_2]$, the resulting state is $(\I,G^\dag\I,\I,G^\dag\I)$ by applying the rewrite rules. 
    Namely, both classes can be written as $(\I,M\I,\I,M\I)$ for $M\in\{A,G^\dag\}$. 
    
    We prove the theorem by induction on the symbolic execution of each abstract gate:
    \begin{enumerate}
        \item Base case: The initial state is $(I,0,0,0)\in (\I,M\I,\I,M\I)$. 

        \item $CX[q_1,q_0]$: the action literally swaps the first and the third component, so it acts trivially on $(\I,M\I,\I,M\I)$. 

\item $W[q_0]$: this gate acts as 
        \begin{align}
            \begin{bmatrix}
                W_{00} & 0 & W_{01} & 0 \\
                0 & W_{00} & 0 & W_{01} \\
                W_{10} & 0 & W_{11} & 0 \\
                0 & W_{10} & 0 & W_{11} \\
            \end{bmatrix}
            \begin{bmatrix}
        \I \\ M\I \\ \I \\ M\I
    \end{bmatrix}
    = 
    \begin{bmatrix}
        W_{00}\I+W_{01}\I \\ W_{00}M\I+W_{01}M\I \\ W_{10}\I+W_{11}\I \\ W_{10}M\I+W_{11}M\I
    \end{bmatrix}
    = 
    \begin{bmatrix}
        \I \\ M\I \\ \I \\ M\I
    \end{bmatrix}.
        \end{align}
    The last equality holds since each entry $W_{ij}$ is a complex number, which may be absorbed into the equivalent class. 
    The transition $M\I\pm M\I\to M\I$ is not directly used in our implementation; the merge in \eq{u-application} is only included to simplify the presentation. \label{itm:w-complete}

        \item $U[q_1,q_2]$: applying rewrite rules to the symbolic execution gives 
        \begin{align}\label{eq:17}
            \begin{bmatrix}
        H & G & 0 & 0 \\
        A & C & 0 & 0 \\
        0 & 0 & H & G \\
        0 & 0 & A & C
    \end{bmatrix}   
    \begin{bmatrix}
        \I \\ G^\dag\I \\ \I \\ G^\dag\I
    \end{bmatrix}
    = 
    \begin{bmatrix}
        \I+GG^\dag\I \\ A\I+CG^\dag\I \\ \I+GG^\dag\I \\ A\I+CG^\dag\I
    \end{bmatrix}
    &\xrightarrow{\mathrm{R}2} 
    \begin{bmatrix}
        \I+(I-HH^\dag)\I \\ A\I+CG^\dag\I \\ \I+(I-HH^\dag)\I \\ A\I+CG^\dag\I
    \end{bmatrix} \\\label{eq:u-application}
    &\xrightarrow{\mathrm{R}6} 
    \begin{bmatrix}
        \I+(I-HH^\dag)\I \\ A\I-AH^\dag\I \\ \I+(I-HH^\dag)\I \\ A\I-AH^\dag\I
    \end{bmatrix}
= 
    \begin{bmatrix}
        \I \\ A\I \\ \I \\ A\I
    \end{bmatrix} .
        \end{align}
    By definition, $H\I=\I$. 
    \label{itm:u-complete}

    \item $U^\dag[q_1,q_2]$: similarly as in \itm{u-complete}, applying rewrite rules to the symbolic execution gives
    \begin{align}
        \begin{bmatrix}
        H^\dag & A^\dag & 0 & 0 \\
        G^\dag & C^\dag & 0 & 0 \\
        0 & 0 & H^\dag & A^\dag \\
        0 & 0 & G^\dag & C^\dag
    \end{bmatrix}   
    \begin{bmatrix}
        \I \\ A\I \\ \I \\ A\I
    \end{bmatrix}
    = 
    \begin{bmatrix}
        \I+A^\dag A\I \\ G^\dag\I+C^\dag A\I \\ \I+A^\dag A\I \\ G^\dag\I+C^\dag A\I \\
    \end{bmatrix}
    &\xrightarrow{\mathrm{R}1}
    \begin{bmatrix}
        \I+(I-H^\dag H)\I \\ G^\dag\I+C^\dag A\I \\ \I+(I-H^\dag H)\I \\ G^\dag\I+C^\dag A\I \\
    \end{bmatrix} \\\label{eq:20}
    &\xrightarrow{\mathrm{R}5}
    \begin{bmatrix}
        \I+(I-H^\dag H)\I \\ G^\dag\I-G^\dag H\I \\ \I+(I-H^\dag H)\I \\ G^\dag\I-G^\dag H\I \\
    \end{bmatrix} 
    =
    \begin{bmatrix}
        \I \\ G^\dag\I \\ \I \\ G^\dag\I
    \end{bmatrix}.
    \end{align}
    \end{enumerate}
    To save space, we have omitted the application of the rules $\mathrm S4$, and $\mathrm R4$ because they are straightforward.

    Since rewriting is applied after each gate, every monomial in \eq{17}--\eq{20} contains at most one adjacent pair of non-principal blocks, so at most one of R1–R8 applies and the derivation shown is the one \alg{extract-polynomial}  performs.
\end{proof}

\dfn{well-formed} describes the standard single-phase-ancilla QSP sequence. The proof extends unchanged to multiple phase ancillas with arbitrary gates on the phase register $q_0$, and to phases conditioned on the block-encoding ancilla being all-zero, because both act on each (principal, non-principal) pair by scalar coefficients, which is all that \itm{w-complete} uses.

\section{Verification of Polynomial Approximation}\label{sec:qsp-verification}

Following the symbolic execution of QSP, when the circuit is programmed correctly, we obtain a block encoding of $p(H)$ for some polynomial $p$. 
To conclude that it block encodes a polynomial approximation of $e^{-\i H t}$ for the given time $t$, we check the worst-case closeness between the two functions in the interval $[-1,1]$, i.e., if there exists a complex subnormalization factor $\beta_*$ such that 
\begin{align}\label{eq:close-infinity-norm}
    \max_{x\in[-1,1]} |e^{-\i x t}-\beta_* \cdot p(x)| \leq \varepsilon.
\end{align}
In standard algorithm designs, this subnormalization parameter $\beta_*$ may be further eliminated by applying an oblivious amplitude amplification. 
In this work, we focus on the approximation of a subnormalized approximation of $e^{-\i H t}$.

To see why it suffices to show that two complex-valued functions are close, we note that $p(H)$ and $e^{-\i H t}$ are diagonalized simultaneously in the eigenbasis of $H$, whose spectrum is contained in $[-1,1]$.
Therefore the spectral norm of $e^{-\i H t}- \beta_*\cdot p(H)$ is bounded by 
the maximum pointwise error between the two functions in the same interval \cite{10.1145/3313276.3316366}.

To obtain a correct Hamiltonian simulation algorithm, \citet{10.1145/3313276.3316366} and subsequent works take a truncation of $\cos(Ht)$ or $\sin(Ht)$, both expanded using the Jacobi-Anger expansion, and then use an algorithm to find the corresponding angles.
While this particular choice of polynomial offer an optimal low-degree approximation of the exponential function, it is not the only polynomial that satisfies (\ref{eq:close-infinity-norm}). 
Therefore, for verification purposes, we offer flexibility on the choices of the polynomials by directly checking that the sufficient condition in (\ref{eq:close-infinity-norm}) is satisfied. 

To verify the polynomial satisfies (\ref{eq:close-infinity-norm}), we take Chebyshev nodes 
\begin{align}\label{eq:chebyshev-nodes}
\left\{x_m=\cos\left(\frac{2m-1}{2M}\pi\right): m\in [M]\right\}
\end{align}
and sample the functions at these points.
The verifier chooses a candidate normalization factor by least-squares
fitting at these nodes, then checks its approximation error.

For $x_m$ defined in \eq{chebyshev-nodes},
let ${\bar e} =(e^{-\i x_m t}: m\in[M])$ and $\bar p=(p(x_m): m\in [M])$ be the sampling of these functions at the grid points.
For $\bar p\neq 0$, \lem{minimization} gives the least-squares choice
$\beta_*=\langle\bar p,\bar e\rangle/\|\bar p\|^2$.
This minimizes $\|\beta\bar p-\bar e\|_2$, not the maximum error
over $[-1,1]$. The guarantee comes from the residual checks below
and the subsequent grid-to-continuum bound, not from optimality
of the chosen factor.
Then the verifier checks whether at each $x_m$, the evaluation of both functions are $(\epsilon/2)$-close, i.e., 
\(
    | \bar e_m
    - \beta_*\cdot \bar p_m| \leq \epsilon/2 \text{ for every $m\in [M] $}.
\)
where the parameter $\beta*$ can be found via an inner product between the two discretizations:
\(
    \beta_* = \langle \bar p,\bar e\rangle/\|\bar p\|^2
\)
by invoking \lem{minimization}.
The following theorem shows that it suffices to evaluate the function at $M=4\max\{D,d\}$ points for $D=\lceil e|t|/2+\ln(32/\epsilon)\rceil$ and $d=\deg(p)$.
\begin{theorem}\label{thm:lipschitz-chebyshev}
    For $\epsilon>0$, let $g$ be a function that can be approximated by a degree-$D$ polynomial to error within $\epsilon/8$ on $[-1,1]$, and $p$ be a degree-$d$ polynomial, 
and
    \(
    x_m=\cos\left(\frac{2m-1}{2M}\pi\right)\text{ for } m\in[M].
    \)
    For $M\geq 4\max\{D,d\}$, if $|g(x_m)-p(x_m)|\leq\epsilon/3$ for all $m$, then $|g(x)-p(x)|\leq\epsilon$ on $[-1,1]$.
\end{theorem}
\begin{proof}
    Let $q$ be the degree-$D$ polynomial that approximates $g$ to error within $\epsilon/8$. Since $q-p$ is a degree-$\max\{D,d\}$ polynomial, by the Ehlich-Zeller theorem (see \citet{ehlichSchwankungPolynomenZwischen1964} and \citet[Theorem~2]{bos2012low}), 
    \[
        \max_{x\in[-1,1]}|q(x)-p(x)| \leq \frac{1}{\cos(\pi/8)} \max_{m\in [M]}|q(x_m)-p(x_m)| \leq \frac{5\sqrt 2 \epsilon}{12\cos(\pi/8)}.
    \]
    By triangle inequality,
    \[
        \max_{x\in[-1,1]}|g(x)-p(x)| \leq \frac{5\epsilon}{12\cos(\pi/8)} + \frac{\epsilon}{8} < \epsilon.
    \]
\end{proof}

Recall that in the previous stage, the verifier performs a symbolic execution to obtain a polynomial $p$.
Then it evaluates both the target function and the polynomial at every $x_m$. 
By Theorem~\ref{thm:lipschitz-chebyshev}, if they are at most $(\epsilon/2)$-close at these points, we can guarantee that the functions must be $\epsilon$-close in $[-1,1]$. 
We then summarize the verifier's action in Algorithm~\ref{alg:verify-hamiltonian-simulation}, and conclude its soundness.

\begin{corollary}\label{cor:soundness-polynomial-approximation}
    For $f\in\{e^{-\i t(\cdot)},\cos(t(\cdot)),\sin(t(\cdot))\}$, \alg{verify-polynomial-approximation} is sound. 
\end{corollary}
\begin{proof}
    We know that the acceptance of the algorithm implies that for every $m\in[M]$, $|f(x_m)-{\beta^*}p(x_m)|\leq\epsilon/2$.
    Since for $D=\lceil e|t|/2+\ln(32/\epsilon)\rceil$, the degree-$D$ truncation of the Jacobi-Anger expansion approximates $f$ to error within $\epsilon/8$. 
    Then by Theorem~\ref{thm:lipschitz-chebyshev}, this implies that $|f(x)-{\beta^*}p(x)|\leq \epsilon$. 
\end{proof}

\begin{algorithm}[t]
\DontPrintSemicolon
\caption{$\VerifyPolynomialApproximation(p,f,t,\epsilon)$}
\label{alg:verify-polynomial-approximation}

\KwIn{
Polynomial $p(x)$; target function $f(x)$;  
time $t$;
tolerance $\varepsilon$
}
\KwOut{
$(\Accept,\beta)$ if $\beta\cdot p$ approximates $f$ to error within
$\epsilon$; otherwise $(\Reject,\bot)$.
}

Initialize $D\gets \lceil e|t|/2 + \ln(32/\epsilon)\rceil$; Set $d$ to be the degree of $p$; Set $M\gets 4\max\{D,d\}$;

Initialize $\bar x\gets (0^M)$; $\bar p\gets (0^M)$; $\bar e\gets (0^M)$;

\ForEach{$m$ in $[M]$}{
    Set $\bar x_m\gets\cos\Big(\frac{2m-1}{2M} \pi\Big)$; Set $\bar e_m\gets f(x_m)$
; Set $\bar p_m \gets p(x_m)$;
}
\lIf{$\|\bar p\|^2\leq 0$}{\Return $(\Reject,\bot)$}; 
Set $\beta\gets \langle\bar p,\bar e\rangle/\|\bar p\|^2$;

\ForEach{$m$ in $[M]$}{

    \lIf{$|f(x_m)-\beta\cdot p(x_m)|>\varepsilon/2$}{\Return  $(\Reject,\bot)$}
}
\Return $(\Accept,\beta)$
\end{algorithm}

\subsection{Example: Verifying the Cosine Component of Hamiltonian Simulation by QSP}

The following example proves the correctness of the cosine component of Hamiltonian simulation based on QSP.

\begin{example}
For simplicity and demonstration purposes in this section, we choose a large $\epsilon=4\%$ and a small evolution time $t=2$ for $H=\frac{X-Z}{2}$, and only use a degree-$4$ approximation of the cosine function by applying a truncation to the Jacobi-Anger expansion, as in Example~\ref{ex:cosine-truncation}.
Since the circuit complexity may be reduced if the angles are represented in low precision, we truncate the angles in (\ref{eq:angles}) to two decimal places, yielding 
$(\phi_0,\phi_1,\phi_2,\phi_3,\phi_4)
    = (2.30, -1.05, 1.07, -1.05, 0.73)$.
We then run a symbolic execution to obtain the polynomial from these truncated angles, listed in Table~\ref{tab:qsvt-symbolic-execution}.
Here to make the symbolic execution easier to follow, we keep the phase factors of the form $e^{\i\phi_j}$ until the end of the execution. 

The next step is to simplify the symbolic expression to a polynomial in $H$:
\[
    e^{-\ii 0.32}\, H^{\dagger}\bigl(H\Sabbr + G\Tabbr\bigr) + e^{\ii 1.78}\, A^{\dagger}\bigl(A\Sabbr + C\Tabbr\bigr),
\]
where $\Sabbr \coloneqq e^{\ii 2.32}\,H^{\dagger}H + e^{\ii 4.42}\,A^{\dagger}A$
and
$\Tabbr \coloneqq e^{\ii 0.18}\,G^{\dagger}H + e^{\ii 2.28}\,C^{\dagger}A$.
Applying the rules and take the real part of the resulting polynomial, we have 
\begin{align}\label{eq:approximate-p}
    p(x) \approx 0.588010 x^4 - 2.000699 x^2 + 0.996542.
\end{align}
Taking the real part may be implemented by applying a linear combination of unitary, but here for exposition, we omit the detail for implementing this particular operation.
We observe that the polynomial (\ref{eq:approximate-p}) has coefficients close to those of (\ref{eq:approximate-p-high-precision}), since the former is obtained by applying a truncation to the angles for the latter.

We then can numerically evaluate both functions at the grid points based on Theorem~\ref{thm:lipschitz-chebyshev}.
To meet the specification $\epsilon=4\%$, we evaluate the function at the Chebyshev nodes defined in \eq{chebyshev-nodes}. 
The deviation is visualized in Fig.~\ref{fig:numerical-difference}. \qed

\begin{figure}
    \centering
    \begin{tikzpicture}
\begin{axis}[
  width=12cm, height=7cm,
  xlabel={$x$},
  ylabel={$p(x) - \cos(2x)\;[10^{-3}]$},
  xmin=-1.02, xmax=1.02,
  ymin=-25, ymax=25,
  axis lines=left,
  legend style={draw=none, fill=none, at={(0.5,1.02)}, anchor=south, legend columns=2, /tikz/every even column/.append style={column sep=1em}},
  tick label style={font=\small}, label style={font=\small},
]
\addplot[black!60, dashed, very thick, domain=-1.02:1.02] {20};
\addlegendentry{$\pm\epsilon/2$}
\addplot[black!60, dashed, very thick, domain=-1.02:1.02, forget plot] {-20};
\addplot[blue!70!black, very thick] coordinates {
(-1.00000,-0.00016)
(-0.99500,-0.82911)
(-0.99000,-1.62416)
(-0.98500,-2.38613)
(-0.98000,-3.11585)
(-0.97500,-3.81414)
(-0.97000,-4.48178)
(-0.96500,-5.11957)
(-0.96000,-5.72828)
(-0.95500,-6.30866)
(-0.95000,-6.86146)
(-0.94500,-7.38742)
(-0.94000,-7.88727)
(-0.93500,-8.36171)
(-0.93000,-8.81144)
(-0.92500,-9.23716)
(-0.92000,-9.63953)
(-0.91500,-10.01923)
(-0.91000,-10.37690)
(-0.90500,-10.71319)
(-0.90000,-11.02873)
(-0.89500,-11.32414)
(-0.89000,-11.60002)
(-0.88500,-11.85698)
(-0.88000,-12.09560)
(-0.87500,-12.31645)
(-0.87000,-12.52010)
(-0.86500,-12.70711)
(-0.86000,-12.87801)
(-0.85500,-13.03335)
(-0.85000,-13.17364)
(-0.84500,-13.29940)
(-0.84000,-13.41112)
(-0.83500,-13.50931)
(-0.83000,-13.59445)
(-0.82500,-13.66701)
(-0.82000,-13.72744)
(-0.81500,-13.77622)
(-0.81000,-13.81377)
(-0.80500,-13.84053)
(-0.80000,-13.85694)
(-0.79500,-13.86340)
(-0.79000,-13.86033)
(-0.78500,-13.84812)
(-0.78000,-13.82716)
(-0.77500,-13.79783)
(-0.77000,-13.76050)
(-0.76500,-13.71554)
(-0.76000,-13.66330)
(-0.75500,-13.60412)
(-0.75000,-13.53835)
(-0.74500,-13.46631)
(-0.74000,-13.38833)
(-0.73500,-13.30472)
(-0.73000,-13.21578)
(-0.72500,-13.12181)
(-0.72000,-13.02311)
(-0.71500,-12.91995)
(-0.71000,-12.81261)
(-0.70500,-12.70136)
(-0.70000,-12.58645)
(-0.69500,-12.46815)
(-0.69000,-12.34669)
(-0.68500,-12.22231)
(-0.68000,-12.09525)
(-0.67500,-11.96574)
(-0.67000,-11.83398)
(-0.66500,-11.70020)
(-0.66000,-11.56459)
(-0.65500,-11.42736)
(-0.65000,-11.28870)
(-0.64500,-11.14879)
(-0.64000,-11.00781)
(-0.63500,-10.86595)
(-0.63000,-10.72336)
(-0.62500,-10.58020)
(-0.62000,-10.43665)
(-0.61500,-10.29284)
(-0.61000,-10.14891)
(-0.60500,-10.00502)
(-0.60000,-9.86130)
(-0.59500,-9.71787)
(-0.59000,-9.57485)
(-0.58500,-9.43237)
(-0.58000,-9.29054)
(-0.57500,-9.14947)
(-0.57000,-9.00927)
(-0.56500,-8.87002)
(-0.56000,-8.73183)
(-0.55500,-8.59479)
(-0.55000,-8.45898)
(-0.54500,-8.32448)
(-0.54000,-8.19137)
(-0.53500,-8.05973)
(-0.53000,-7.92961)
(-0.52500,-7.80110)
(-0.52000,-7.67423)
(-0.51500,-7.54909)
(-0.51000,-7.42570)
(-0.50500,-7.30414)
(-0.50000,-7.18443)
(-0.49500,-7.06663)
(-0.49000,-6.95077)
(-0.48500,-6.83689)
(-0.48000,-6.72502)
(-0.47500,-6.61519)
(-0.47000,-6.50742)
(-0.46500,-6.40174)
(-0.46000,-6.29818)
(-0.45500,-6.19673)
(-0.45000,-6.09743)
(-0.44500,-6.00028)
(-0.44000,-5.90529)
(-0.43500,-5.81247)
(-0.43000,-5.72182)
(-0.42500,-5.63335)
(-0.42000,-5.54705)
(-0.41500,-5.46292)
(-0.41000,-5.38095)
(-0.40500,-5.30115)
(-0.40000,-5.22349)
(-0.39500,-5.14798)
(-0.39000,-5.07459)
(-0.38500,-5.00332)
(-0.38000,-4.93414)
(-0.37500,-4.86704)
(-0.37000,-4.80200)
(-0.36500,-4.73900)
(-0.36000,-4.67801)
(-0.35500,-4.61902)
(-0.35000,-4.56199)
(-0.34500,-4.50690)
(-0.34000,-4.45373)
(-0.33500,-4.40245)
(-0.33000,-4.35302)
(-0.32500,-4.30541)
(-0.32000,-4.25960)
(-0.31500,-4.21555)
(-0.31000,-4.17323)
(-0.30500,-4.13261)
(-0.30000,-4.09364)
(-0.29500,-4.05630)
(-0.29000,-4.02055)
(-0.28500,-3.98636)
(-0.28000,-3.95367)
(-0.27500,-3.92247)
(-0.27000,-3.89271)
(-0.26500,-3.86436)
(-0.26000,-3.83737)
(-0.25500,-3.81171)
(-0.25000,-3.78734)
(-0.24500,-3.76422)
(-0.24000,-3.74231)
(-0.23500,-3.72158)
(-0.23000,-3.70198)
(-0.22500,-3.68348)
(-0.22000,-3.66605)
(-0.21500,-3.64963)
(-0.21000,-3.63420)
(-0.20500,-3.61971)
(-0.20000,-3.60614)
(-0.19500,-3.59344)
(-0.19000,-3.58157)
(-0.18500,-3.57050)
(-0.18000,-3.56020)
(-0.17500,-3.55063)
(-0.17000,-3.54175)
(-0.16500,-3.53354)
(-0.16000,-3.52595)
(-0.15500,-3.51896)
(-0.15000,-3.51254)
(-0.14500,-3.50664)
(-0.14000,-3.50125)
(-0.13500,-3.49633)
(-0.13000,-3.49185)
(-0.12500,-3.48779)
(-0.12000,-3.48411)
(-0.11500,-3.48080)
(-0.11000,-3.47782)
(-0.10500,-3.47515)
(-0.10000,-3.47277)
(-0.09500,-3.47065)
(-0.09000,-3.46878)
(-0.08500,-3.46712)
(-0.08000,-3.46567)
(-0.07500,-3.46440)
(-0.07000,-3.46330)
(-0.06500,-3.46235)
(-0.06000,-3.46153)
(-0.05500,-3.46083)
(-0.05000,-3.46024)
(-0.04500,-3.45974)
(-0.04000,-3.45932)
(-0.03500,-3.45897)
(-0.03000,-3.45869)
(-0.02500,-3.45847)
(-0.02000,-3.45829)
(-0.01500,-3.45816)
(-0.01000,-3.45807)
(-0.00500,-3.45802)
(0.00000,-3.45800)
(0.00500,-3.45802)
(0.01000,-3.45807)
(0.01500,-3.45816)
(0.02000,-3.45829)
(0.02500,-3.45847)
(0.03000,-3.45869)
(0.03500,-3.45897)
(0.04000,-3.45932)
(0.04500,-3.45974)
(0.05000,-3.46024)
(0.05500,-3.46083)
(0.06000,-3.46153)
(0.06500,-3.46235)
(0.07000,-3.46330)
(0.07500,-3.46440)
(0.08000,-3.46567)
(0.08500,-3.46712)
(0.09000,-3.46878)
(0.09500,-3.47065)
(0.10000,-3.47277)
(0.10500,-3.47515)
(0.11000,-3.47782)
(0.11500,-3.48080)
(0.12000,-3.48411)
(0.12500,-3.48779)
(0.13000,-3.49185)
(0.13500,-3.49633)
(0.14000,-3.50125)
(0.14500,-3.50664)
(0.15000,-3.51254)
(0.15500,-3.51896)
(0.16000,-3.52595)
(0.16500,-3.53354)
(0.17000,-3.54175)
(0.17500,-3.55063)
(0.18000,-3.56020)
(0.18500,-3.57050)
(0.19000,-3.58157)
(0.19500,-3.59344)
(0.20000,-3.60614)
(0.20500,-3.61971)
(0.21000,-3.63420)
(0.21500,-3.64963)
(0.22000,-3.66605)
(0.22500,-3.68348)
(0.23000,-3.70198)
(0.23500,-3.72158)
(0.24000,-3.74231)
(0.24500,-3.76422)
(0.25000,-3.78734)
(0.25500,-3.81171)
(0.26000,-3.83737)
(0.26500,-3.86436)
(0.27000,-3.89271)
(0.27500,-3.92247)
(0.28000,-3.95367)
(0.28500,-3.98636)
(0.29000,-4.02055)
(0.29500,-4.05630)
(0.30000,-4.09364)
(0.30500,-4.13261)
(0.31000,-4.17323)
(0.31500,-4.21555)
(0.32000,-4.25960)
(0.32500,-4.30541)
(0.33000,-4.35302)
(0.33500,-4.40245)
(0.34000,-4.45373)
(0.34500,-4.50690)
(0.35000,-4.56199)
(0.35500,-4.61902)
(0.36000,-4.67801)
(0.36500,-4.73900)
(0.37000,-4.80200)
(0.37500,-4.86704)
(0.38000,-4.93414)
(0.38500,-5.00332)
(0.39000,-5.07459)
(0.39500,-5.14798)
(0.40000,-5.22349)
(0.40500,-5.30115)
(0.41000,-5.38095)
(0.41500,-5.46292)
(0.42000,-5.54705)
(0.42500,-5.63335)
(0.43000,-5.72182)
(0.43500,-5.81247)
(0.44000,-5.90529)
(0.44500,-6.00028)
(0.45000,-6.09743)
(0.45500,-6.19673)
(0.46000,-6.29818)
(0.46500,-6.40174)
(0.47000,-6.50742)
(0.47500,-6.61519)
(0.48000,-6.72502)
(0.48500,-6.83689)
(0.49000,-6.95077)
(0.49500,-7.06663)
(0.50000,-7.18443)
(0.50500,-7.30414)
(0.51000,-7.42570)
(0.51500,-7.54909)
(0.52000,-7.67423)
(0.52500,-7.80110)
(0.53000,-7.92961)
(0.53500,-8.05973)
(0.54000,-8.19137)
(0.54500,-8.32448)
(0.55000,-8.45898)
(0.55500,-8.59479)
(0.56000,-8.73183)
(0.56500,-8.87002)
(0.57000,-9.00927)
(0.57500,-9.14947)
(0.58000,-9.29054)
(0.58500,-9.43237)
(0.59000,-9.57485)
(0.59500,-9.71787)
(0.60000,-9.86130)
(0.60500,-10.00502)
(0.61000,-10.14891)
(0.61500,-10.29284)
(0.62000,-10.43665)
(0.62500,-10.58020)
(0.63000,-10.72336)
(0.63500,-10.86595)
(0.64000,-11.00781)
(0.64500,-11.14879)
(0.65000,-11.28870)
(0.65500,-11.42736)
(0.66000,-11.56459)
(0.66500,-11.70020)
(0.67000,-11.83398)
(0.67500,-11.96574)
(0.68000,-12.09525)
(0.68500,-12.22231)
(0.69000,-12.34669)
(0.69500,-12.46815)
(0.70000,-12.58645)
(0.70500,-12.70136)
(0.71000,-12.81261)
(0.71500,-12.91995)
(0.72000,-13.02311)
(0.72500,-13.12181)
(0.73000,-13.21578)
(0.73500,-13.30472)
(0.74000,-13.38833)
(0.74500,-13.46631)
(0.75000,-13.53835)
(0.75500,-13.60412)
(0.76000,-13.66330)
(0.76500,-13.71554)
(0.77000,-13.76050)
(0.77500,-13.79783)
(0.78000,-13.82716)
(0.78500,-13.84812)
(0.79000,-13.86033)
(0.79500,-13.86340)
(0.80000,-13.85694)
(0.80500,-13.84053)
(0.81000,-13.81377)
(0.81500,-13.77622)
(0.82000,-13.72744)
(0.82500,-13.66701)
(0.83000,-13.59445)
(0.83500,-13.50931)
(0.84000,-13.41112)
(0.84500,-13.29940)
(0.85000,-13.17364)
(0.85500,-13.03335)
(0.86000,-12.87801)
(0.86500,-12.70711)
(0.87000,-12.52010)
(0.87500,-12.31645)
(0.88000,-12.09560)
(0.88500,-11.85698)
(0.89000,-11.60002)
(0.89500,-11.32414)
(0.90000,-11.02873)
(0.90500,-10.71319)
(0.91000,-10.37690)
(0.91500,-10.01923)
(0.92000,-9.63953)
(0.92500,-9.23716)
(0.93000,-8.81144)
(0.93500,-8.36171)
(0.94000,-7.88727)
(0.94500,-7.38742)
(0.95000,-6.86146)
(0.95500,-6.30866)
(0.96000,-5.72828)
(0.96500,-5.11957)
(0.97000,-4.48178)
(0.97500,-3.81414)
(0.98000,-3.11585)
(0.98500,-2.38613)
(0.99000,-1.62416)
(0.99500,-0.82911)
(1.00000,-0.00016)
};
\addlegendentry{$p(x) - \cos(2x)$}
\end{axis}
\end{tikzpicture}
    \caption{Deviation of $p(x)$ from $\cos(2x)$ for $x\in[-1,1]$, where $p$ is defined in (\ref{eq:approximate-p}).}
    \label{fig:numerical-difference}
\end{figure}

\begin{table*}[t]
\centering

\resizebox{\textwidth}{!}{\begin{tabular}{l|cccc}
\toprule
$op$ & $00$ & $01$ & $10$ & $11$ \\
\midrule
 & $I$ & $0$ & $0$ & $0$ \\
$CX$ & $0$ & $0$ & $I$ & $0$ \\
$R_z(2\psi_0)$ & $0$ & $0$ & $e^{\ii 2.30} I$ & $0$ \\
$CX$ & $e^{\ii 2.30} I$ & $0$ & $0$ & $0$ \\\midrule
$U_H$ & $e^{\ii 2.30} H$ & $e^{\ii 2.30} A$ & $0$ & $0$ \\
\midrule
$CX$ & $0$ & $e^{\ii 2.30} A$ & $e^{\ii 2.30} H$ & $0$ \\
$R_z(2\phi_1)$ & $0$ & $e^{\ii 3.35} A$ & $e^{\ii 1.25} H$ & $0$ \\
$CX$ & $e^{\ii 1.25} H$ & $e^{\ii 3.35} A$ & $0$ & $0$ \\\midrule
$(U_H)^{\dagger}$ & $e^{\ii 1.25} H^{\dagger}H + e^{\ii 3.35} A^{\dagger}A$ & $e^{\ii 1.25} G^{\dagger}H + e^{\ii 3.35} C^{\dagger}A$ & $0$ & $0$ \\
\midrule
$CX$ & $0$ & $e^{\ii 1.25} G^{\dagger}H + e^{\ii 3.35} C^{\dagger}A$ & $e^{\ii 1.25} H^{\dagger}H + e^{\ii 3.35} A^{\dagger}A$ & $0$ \\
$R_z(2\phi_2)$ & $0$ & $\underbrace{e^{\ii 0.18}\, G^{\dagger}H + e^{\ii 2.28}\, C^{\dagger}A}_{\eqqcolon\;\Tabbr}$ & $\underbrace{e^{\ii 2.32}\, H^{\dagger}H + e^{\ii 4.42}\, A^{\dagger}A}_{\eqqcolon\;\Sabbr}$ & $0$ \\
$CX$ & $\Sabbr$ & $\Tabbr$ & $0$ & $0$ \\\midrule
$U_H$ & $H\Sabbr + G\Tabbr$ & $A\Sabbr + C\Tabbr$ & $0$ & $0$ \\
\midrule
$CX$ & $0$ & $A\Sabbr + C\Tabbr$ & $H\Sabbr + G\Tabbr$ & $0$ \\
$R_z(2\phi_3)$ & $0$ & $e^{\ii 1.05}\,\bigl(A\Sabbr + C\Tabbr\bigr)$ & $e^{-\ii 1.05}\,\bigl(H\Sabbr + G\Tabbr\bigr)$ & $0$ \\
$CX$ & $e^{-\ii 1.05}\,\bigl(H\Sabbr + G\Tabbr\bigr)$ & $e^{\ii 1.05}\,\bigl(A\Sabbr + C\Tabbr\bigr)$ & $0$ & $0$ \\\midrule
$(U_H)^{\dagger}$ & $e^{-\ii 1.05} H^{\dagger}\bigl(H\Sabbr + G\Tabbr\bigr) + e^{\ii 1.05} A^{\dagger}\bigl(A\Sabbr + C\Tabbr\bigr)$ & $e^{-\ii 1.05} G^{\dagger}\bigl(H\Sabbr + G\Tabbr\bigr) + e^{\ii 1.05} C^{\dagger}\bigl(A\Sabbr + C\Tabbr\bigr)$ & $0$ & $0$ \\
\midrule
$CX$ & $0$ & $e^{-\ii 1.05} G^{\dagger}\bigl(H\Sabbr + G\Tabbr\bigr) + e^{\ii 1.05} C^{\dagger}\bigl(A\Sabbr + C\Tabbr\bigr)$ & $e^{-\ii 1.05} H^{\dagger}\bigl(H\Sabbr + G\Tabbr\bigr) + e^{\ii 1.05} A^{\dagger}\bigl(A\Sabbr + C\Tabbr\bigr)$ & $0$ \\
$R_z(2\phi_4)$ & $0$ & $e^{-\ii 1.78}\, G^{\dagger}\bigl(H\Sabbr + G\Tabbr\bigr) + e^{\ii 0.32}\, C^{\dagger}\bigl(A\Sabbr + C\Tabbr\bigr)$ & $e^{-\ii 0.32}\, H^{\dagger}\bigl(H\Sabbr + G\Tabbr\bigr) + e^{\ii 1.78}\, A^{\dagger}\bigl(A\Sabbr + C\Tabbr\bigr)$ & $0$ \\
$CX$ & $e^{-\ii 0.32}\, H^{\dagger}\bigl(H\Sabbr + G\Tabbr\bigr) + e^{\ii 1.78}\, A^{\dagger}\bigl(A\Sabbr + C\Tabbr\bigr)$ & $e^{-\ii 1.78}\, G^{\dagger}\bigl(H\Sabbr + G\Tabbr\bigr) + e^{\ii 0.32}\, C^{\dagger}\bigl(A\Sabbr + C\Tabbr\bigr)$ & $0$ & $0$ \\
\bottomrule
\end{tabular}}
\caption{Symbolic execution of the degree-4 QSP circuit with phases
$\phi = (2.30,\, -1.05,\, 1.07,\, -1.05,\, 0.73)$.
Shorthands:
$\Sabbr \coloneqq e^{\ii 2.32}\,H^{\dagger}H + e^{\ii 4.42}\,A^{\dagger}A$
and
$\Tabbr \coloneqq e^{\ii 0.18}\,G^{\dagger}H + e^{\ii 2.28}\,C^{\dagger}A$.}
\label{tab:qsvt-symbolic-execution}
\end{table*}
\end{example}

\section{Overall Verification Algorithm}
\label{sec:overall-algorithm}

We now summarize the full verification procedure. The verifier takes as input a
target Hamiltonian $H$, an alleged block-encoding circuit $C_{\mathrm{BE}}$, a
QSP circuit $C_{\mathrm{QSP}}$, an evolution time $t$, and an error tolerance
$\varepsilon$. It proceeds by invoking the four verification stages described
above as subroutines.

\begin{algorithm}[t]
\DontPrintSemicolon
\caption{$\VerifyHamiltonianSimulation(H,C_{\BE},C_{\QSP},t,\epsilon)$}
\label{alg:verify-hamiltonian-simulation}

\KwIn{
Target Hamiltonian $H=\sum_P h_P P$;
block-encoding circuit $C_{\mathrm{BE}}$;
QSP circuit $C_{\mathrm{QSP}}$;
evolution time $t>0$;
tolerance $\varepsilon$
}
\KwOut{
(\Accept, $\beta$) if the QSP circuit implements a
subnormalized $\epsilon$-approximation of $e^{-\i Ht}$ with scaling factor
$\beta$; otherwise output \Reject.
}

$(s_1,\alpha)\gets \textsc{VerifyBlockEncoding}(H,C_{\BE},\frac{\epsilon}{2|t|})$;

\lIf{$s_1 = \Reject$}{
    \Return \Reject
}

$E\gets\textsc{SymbolicExecute}(C_{\mathrm{QSP}})$;

$(s_3,p)\gets
\textsc{ExtractPolynomial}(E)$;

\lIf{$s_3=\Reject$}{\Return \textsc{Reject}}

$(s_{4},\beta)\gets\VerifyPolynomialApproximation\bigl(p, e^{-i\alpha t(\cdot)}, {\alpha t},\epsilon/2\bigr)$;

\Return $(s_4,\beta)$
\end{algorithm}

\subsection{Soundness}\label{subsec:overall-soundness}

In this section, we prove the soundness of \alg{verify-hamiltonian-simulation}, i.e., 
when the algorithm outputs \textsc{Accept}, the resulting quantum circuit is an $\epsilon$-approximation of $e^{-\i H t}$ up to a subnormalization factor. 

\begin{theorem}[Soundness of \alg{verify-hamiltonian-simulation}]
If \alg{verify-hamiltonian-simulation} returns $(\textsc{Accept},\beta)$, then the principal block
$B_{\mathrm{QSP}}$ of the QSP circuit satisfies
$\|\beta B_{\mathrm{QSP}}-e^{-\i Ht}\|\leq\epsilon$.
\end{theorem}
\begin{proof}
    When the algorithm accepts, it must be true that $s_1,s_3,s_4$ are all \Accept. 
    By the soundness statement in \thm{soundness-verify-block-encoding}, \thm{qsp-symbolic-execution}, \cor{soundness-extract-polynomial}, and \cor{soundness-polynomial-approximation}, it remains to show that the total error is within $\epsilon$. 
    By the soundness of \alg{verify-block-encoding}, we know that the all-zero branch has matrix $B_0$ satisfying
    $\|\alpha B_0 - H\| \leq \frac{\epsilon}{2|t|}$.
    This implies that $B_0$ is $\frac{\epsilon}{2\alpha|t|}$-close to $H/\alpha$.
    By the Lipschitz continuity of the exponential function,  
    $\|e^{-\i B_0 \alpha t}-e^{-\i H t}\| \leq |t|\cdot \|\alpha B_0- H\| \leq \frac{\epsilon}{2}$.
    
    Then by \cor{soundness-extract-polynomial}, the QSP circuit $C_{\QSP}$ is certified to produce a polynomial $p$.
    Furthermore, by \cor{soundness-polynomial-approximation}, the polynomial $p$ satisfies 
    \(
        \|\beta\cdot p(B_0) - e^{-\i B_0 \alpha t}\| \leq \frac{\epsilon}{2}.
    \)
    Finally, applying triangle inequality, we have
    \[
        \|\beta\cdot p(B_0) - e^{-\i H t}\| \leq \|\beta\cdot p(B_0) - e^{-\i B_0 \alpha t}\| + \|e^{-\i B_0 \alpha t}-e^{-\i H t}\| 
        \leq \epsilon.
    \]
    This means that when \alg{verify-hamiltonian-simulation} accepts, the entire circuit meets the specification, i.e., it simulates the given Hamiltonian at time $t$ to error within $\epsilon$. 
\end{proof}

\paragraph{Composition of the QSP Wrapper and Block Encoding.}
To compose the QSP wrapper with the concrete block-encoding circuit, we identify the $r$ ancilla qubits and $n$ system qubits of the block-encoding circuit with the corresponding block-encoding ancilla and system qubits of the wrapper. The qubits within each register may be matched in any order, provided that the same mapping is used consistently throughout the composed circuit. Each occurrence of $U_H$ or $U_H^\dagger$ is then replaced by the block-encoding circuit $C_{\mathrm{BE}}$ or its adjoint $C_{\mathrm{BE}}^\dagger$, respectively. If an occurrence is controlled by a phase-ancilla qubit, the same control is added to every gate in the corresponding block-encoding circuit. If a gate in the QSP wrapper acts on a phase ancilla and is open-controlled by the abstract block-encoding ancilla, the abstract control is replaced by open controls on all $r$ concrete block-encoding ancilla qubits, while the target remains the same phase ancilla.

\begin{corollary}[Correctness of the composed circuit]\label{cor:composed-circuit}
Let $C_{\mathrm{comp}}$ be the circuit obtained by composing the QSP wrapper
with the verified block-encoding circuit as described above. If
\alg{verify-hamiltonian-simulation} returns $(\Accept,\beta)$, then
$\left\|
\beta(\bra{0^{s+r}}\otimes I)C_{\mathrm{comp}}
(\ket{0^{s+r}}\otimes I)-e^{-\i Ht}
\right\|\leq\epsilon$.
\end{corollary}

\section{Evaluation}

We evaluate the scalability of the verifier on two verification tasks. 

\begin{description}
    \item[TASK1] We measure the size of structured Pauli-sum LCU block encodings that the verifier can handle, reporting runtime and memory as the number of system qubits, Pauli terms, and selector ancillas increase.
    \item[TASK2] We measure the size of QSP Hamiltonian-simulation circuits that the verifier can certify after abstracting the block encoding as an opaque operation. For each circuit, the verifier extracts the all-zero branch and checks the approximation error against \(e^{-\mathrm{i}  x \rmark{\alpha} t}\) on \([-1,1]\).
\end{description}

\subsection{Benchmarks and Circuit Generation}

For TASK1, we use three standard families of 2-local Hamiltonians:
\begin{align*}
H_{\mathrm{Ising}}^{(n)}
&=
\sum_{i=0}^{n-1} Z_i Z_{(i+1)\bmod n}
+
\sum_{i=0}^{n-1} X_i ,
\\
H_{\mathrm{MaxCut}}^{(n)}
&=
\sum_{(i,j)\in E(C_n)}
\frac{I-Z_iZ_j}{2},
\qquad
E(C_n)=\{(i,(i+1)\bmod n)\}_{i=0}^{n-1},
\\
H_{\mathrm{Heisenberg}}^{(n)}
&=
\sum_{i=0}^{n-2}
\left(
X_iX_{i+1}
+
Y_iY_{i+1}
+
Z_iZ_{i+1}
\right).
\end{align*}
For a Hamiltonian \(H=\sum_{\ell=0}^{K-1}c_\ell P_\ell\), the benchmark
generator constructs the standard LCU block encoding
$Q_{\mathrm{BE}}
    =
    (\mathrm{PREPARE}^{\dagger}\otimes I)
    \cdot
    \mathrm{SELECT}
    \cdot
    (\mathrm{PREPARE}\otimes I)$.
Here \(\alpha=\sum_{\ell=0}^{K-1}|c_\ell|\),
\(\mathrm{PREPARE}\) prepares amplitudes proportional to
\(\sqrt{|c_\ell|/\alpha}\), and \(\mathrm{SELECT}\) applies the
corresponding signed Pauli string. In each of our three benchmark
families, the LCU terms have equal absolute coefficients. The
generator therefore prepares equal amplitudes on the \(K\) term
labels and zero amplitude on unused selector states. It synthesizes
\(\mathrm{PREPARE}\) with \(R_y\) rotations controlled by prefixes
of the selector register, and constructs
\(\mathrm{PREPARE}^{\dagger}\) by reversing the gate sequence and
negating the rotation angles.

For TASK2, all benchmarks use a fixed opaque block encoding \(Q_{\mathrm{BE}}\)
with one system qubit and one block-encoding ancilla.  The verifier does not
expand the implementation of \(Q_{\mathrm{BE}}\) or \(Q_{\mathrm{BE}}^\dagger\);
it treats them as symbolic block-encoding operations.  If the principal block of
\(Q_{\mathrm{BE}}\) is \(H/\alpha\), then simulating \(H\) for physical time
\(t\) corresponds to approximating \(e^{-\mathrm{i}\tau x}\), where \(\tau=\alpha t\).  Thus TASK2 is
parameterized by the effective simulation time \(\tau\).

The TASK2 circuits are generated using an optimized QSP construction.  The goal
is to exercise the verifier on circuits produced by a nontrivial
circuit-generation workflow, rather than only on a direct textbook
construction.  A straightforward generator would synthesize two independent QSP
circuits for the cosine and sine components of
$e^{-\mathrm{i}\tau x}
    =
    \cos(\tau x)-\mathrm{i}\sin(\tau x)$.

Our generator instead shares one QSP skeleton between the two components: a
selector qubit multiplexes the cosine and sine phase sequences, and the two
branches share the same positions of \(Q_{\mathrm{BE}}\) and
\(Q_{\mathrm{BE}}^\dagger\) calls.  The extracted cosine and sine responses are
then combined by a two-term LCU.

This optimized construction is not trusted by the verifier.  The verifier is
given only the generated QASM circuit; it symbolically executes the circuit,
extracts the all-zero branch, and checks the polynomial implemented by that
branch.  Thus the benchmark demonstrates a separation between circuit generation
and verification: the generator may use circuit-level QSP optimizations, while
correctness is established by independently checking the generated circuit.

The generator uses \textsc{pyqsp} only as an untrusted source of benchmark phase
sequences and polynomial data.  For each value of \(\tau\), \textsc{pyqsp}
produces bounded Chebyshev approximations
\[
    \widehat P_{\cos}^{(\tau)}(x)
    \approx
    \frac{1}{2}\cos(\tau x),
    \qquad
    \widehat P_{\sin}^{(\tau)}(x)
    \approx
    \frac{1}{2}\sin(\tau x).
\]
The cosine approximation is an even polynomial and the sine approximation is an
odd polynomial, matching the parity structure of standard QSP responses.  The
Chebyshev coefficients are converted to monomial coefficients before
verification.  Further details of the shared-skeleton construction, the
\textsc{pyqsp}-to-QASM phase conversion, and the LCU normalization are given in
Appendix~\ref{app:optimized-hs-construction}.

The all-zero branch extracted by the verifier has the form
$B^{\mathrm{HS}}_{\mathbf 0}(x)
    =
    \frac{1}{2}
    \left(
        \widehat P_{\cos}^{(\tau)}(x)
        -
        \mathrm{i}\widehat P_{\sin}^{(\tau)}(x)
    \right)$.
Since the QSP polynomials produced by the bounded synthesis are themselves
half-scaled, this branch approximates
$B^{\mathrm{HS}}_{\mathbf 0}(x)
    \approx
    \frac{1}{4}e^{-\mathrm{i}\tau x}$.

For approximation checking, the verifier infers a scalar normalization factor
\(\beta\) from the sampled values:
\(
    \beta =
    \langle p,e\rangle/\|p\|^2,
\text{ with entries }
    p_m = B^{\mathrm{HS}}_{\mathbf 0}(x_m)
    \text{ and }
    e_m = e^{-\mathrm{i}\tau x_m}.
\)
It then checks
\[\max_m
    \left|
        \beta B^{\mathrm{HS}}_{\mathbf 0}(x_m)
        - e^{-\mathrm{i}\tau x_m}
    \right|
    \le \epsilon/2.
\]
In our full Hamiltonian-simulation benchmarks, the inferred \(\beta\) is
typically close to \(4\), matching the expected inverse normalization from the
two-term LCU and the half-scaled QSP polynomials.

\subsection{Experimental Setup}

All experiments were run on a single workstation with 8 CPU cores and 16 GiB of RAM, running macOS 13.2 (Darwin 22.3.0, arm64). We use Python 3.11. Each benchmark is executed with a timeout of 1800 seconds.  We report wall-clock verification time and maximum resident set size (RSS).

\paragraph{Numerical accuracy.}
We use the Arb/Acb ball arithmetic provided by Python-FLINT at 60 decimal digits of working precision for all arithmetic computations, including coefficient propagation,
polynomial manipulation, residual evaluation, and error-bound
comparison. To check an inequality $\mathrm{LHS}\leq\mathrm{RHS}$,
the implementation requires the upper bound of the ball enclosing
$\mathrm{LHS}$ to be strictly smaller than the lower bound of the
ball enclosing $\mathrm{RHS}$.

Zero tests are exact. Numerical roundoff may therefore leave
terms whose coefficients would be zero in exact arithmetic.
During simplification, these powers of $H$ are retained rather
than removed on the basis of their small coefficients.

We report \textsc{PASS} when verification establishes the requested
guarantee: the block-encoding error bound for TASK1, as in Theorem~\ref{thm:soundness-verify-block-encoding}, or the full-interval QSP approximation for TASK2, as in Corollary~\ref{cor:soundness-polynomial-approximation}. A \textsc{FAIL} means that the verifier could not establish the guarantee. A \textsc{TIMEOUT} means that verification did not
finish within 1800 seconds.

The implementation applies rewriting after each QSP gate application.

\subsection{TASK1: Block-Encoding Verification}
\label{subsec:eval-rq1}

We evaluate 51 block-encoding benchmarks: 17 periodic Ising instances, 17 cycle
MaxCut instances, and 17 open-chain Heisenberg instances.  The system sizes are
\[
    n\in
    \{2,3,4,5,6,7,8,9,10,16,17,32,33,64,65,128,256\}.
\]
Table~\ref{tab:block-encoding-results} summarizes the block-encoding verification results.
The verifier proves 50 of the 51 instances within the timeout. The only timeout is the largest Heisenberg instance, with $(n=256)$ and 765 Pauli terms. With an extended timeout, the verifier also completes this instance, taking 4976.42 seconds.

\begin{table}[t]
\centering
\footnotesize
\setlength{\tabcolsep}{2.5pt}
\begin{tabular}{lrrrrrrr}
\toprule
Family & \(n\) & Terms & Sel. anc. & Gates &
Time (s) & RSS (MiB) & Result \\
\midrule
Ising & 2   & 4   & 2 & 44       & 0.08   & 58.47  & \textsc{PASS} \\
Ising & 32  & 64  & 6 & 1{,}862  & 3.69   & 62.88  & \textsc{PASS} \\
Ising & 128 & 256 & 8 & 9{,}062  & 155.96 & 101.36 & \textsc{PASS} \\
Ising & 256 & 512 & 9 & 19{,}686 & 807.87 & 257.84 & \textsc{PASS} \\
\midrule
MaxCut & 2   & 4   & 2 & 34       & 0.09   & 58.34  & \textsc{PASS} \\
MaxCut & 32  & 64  & 6 & 1{,}574  & 3.58   & 61.31  & \textsc{PASS} \\
MaxCut & 128 & 256 & 8 & 7{,}654  & 97.92  & 86.39  & \textsc{PASS} \\
MaxCut & 256 & 512 & 9 & 16{,}614 & 575.73 & 169.61 & \textsc{PASS} \\
\midrule
Heisenberg & 2   & 3   & 2  & 38       & 0.08   & 58.12  & \textsc{PASS} \\
Heisenberg & 32  & 93  & 7  & 3{,}384  & 15.43  & 67.75  & \textsc{PASS} \\
Heisenberg & 128 & 381 & 9  & 16{,}152 & 470.81 & 171.19 & \textsc{PASS} \\
Heisenberg & 256 & 765 & 10 & 34{,}712 & 1801.03 & -- & \textsc{TIMEOUT} \\
\bottomrule
\end{tabular}
\caption{Representative block-encoding verification results.
All instances use a tolerance of $\epsilon_{\mathrm{BE}} = 10^{-8}$.}
\label{tab:block-encoding-results}
\end{table}

These results show that the verifier scales with the symbolic branch structure
rather than the Hilbert-space dimension.  The largest successful instance has
256 system qubits and 512 Pauli terms, far beyond the range where constructing
the \(2^n\times 2^n\) system matrix is possible.  The timeout on the largest
Heisenberg instance is consistent with the verifier's cost model: increasing
the number of LCU terms increases the number of symbolic branches and 
Pauli expressions that must be simplified.

In addition, we used our verifier to examine block encodings generated by Cobble (version 0.1.0, commit 6a6b56c) ~\cite{yuan2026cobble} and identified a correctness bug when the number of equally weighted Pauli terms is not a power of two. Cobble's state-preparation routine applies Hadamard gates to the selector ancillas, resulting in nonzero amplitudes for unused selector states. For example, for the two-qubit Heisenberg Hamiltonian $H=X_0X_1+Y_0Y_1+Z_0Z_1$, the generated circuit encodes $(H+I)/4$ instead of the expected $H/3$. This demonstrates that our verifier can detect semantic errors in automatically generated block-encoding circuits.

\subsection{TASK2: Hamiltonian-Simulation QSP Verification}
\label{subsec:eval-rq2}

For each Hamiltonian-simulation benchmark, the verifier symbolically executes
the generated circuit and extracts the all-zero branch polynomial
\(B^{\mathrm{HS}}_{\mathbf{0}}(x)\).  It then applies the polynomial
approximation checker from Section~\ref{sec:qsp-verification}: the checker
infers the normalization factor \(\beta\) on the Chebyshev-root grid and accepts
only if
\(
    \left|
        \beta B^{\mathrm{HS}}_{\mathbf{0}}(x_m)
        -
        e^{-\mathrm{i}\tau x_m}
    \right|
    \le
    \epsilon/2
\)
at every grid point.  By Theorem~\ref{thm:lipschitz-chebyshev}, this finite check certifies
\(
    \max_{x\in[-1,1]}
    \left|
        \beta B^{\mathrm{HS}}_{\mathbf{0}}(x)
        -
        e^{-\mathrm{i}\tau x}
    \right|
    \le
    \epsilon .
\)

\begin{table}[t]
\centering
\footnotesize
\setlength{\tabcolsep}{3pt}
\begin{tabular}{rrrrrrrl}
\toprule
\(\tau\) & \(\epsilon_{\mathrm{QSP}}\) & \(d\) &
Gates & Grid pts. & Time (s) & RSS (MiB) & Result \\
\midrule
0.1  & \(10^{-4}\) & 3  & 113     & 52  & 0.40 & 64.28 & \textsc{PASS} \\
1.0  & \(10^{-4}\) & 7  & 233     & 60  & 0.59 & 62.94 & \textsc{PASS} \\
4.0  & \(10^{-4}\) & 11 & 353     & 76  & 0.43 & 64.23 & \textsc{PASS} \\
10.0 & \(10^{-4}\) & 21 & 653     & 108 & 0.62 & 64.67 & \textsc{PASS} \\
20.0 & \(10^{-4}\) & 35 & 1{,}073 & 160 & 1.01 & 65.34 & \textsc{PASS} \\
50.0 & \(10^{-4}\) & 77 & 2{,}333 & 324 & 3.07 & 66.55 & \textsc{PASS} \\
\midrule
0.5 & \(10^{-1}\)  & 3  & 113 & 28  & 0.35 & 63.61 & \textsc{PASS} \\
0.5 & \(10^{-4}\)  & 5  & 173 & 56  & 0.36 & 63.66 & \textsc{PASS} \\
0.5 & \(10^{-8}\)  & 7  & 233 & 92  & 0.44 & 63.95 & \textsc{PASS} \\
0.5 & \(10^{-12}\) & 11 & 353 & 128 & 0.44 & 64.59 & \textsc{PASS} \\
\bottomrule
\end{tabular}
\caption{Representative TASK2 results. The upper group varies
effective simulation time at fixed
\(\epsilon_{\mathrm{QSP}}=10^{-4}\); the lower group varies
the standalone QSP tolerance at fixed \(\tau=0.5\).}
\label{tab:hamsim-results}
\end{table}

Table~\ref{tab:hamsim-results} summarizes both TASK2 sweeps. All 38
time-sweep circuits and all 12 tolerance-sweep circuits pass the
approximation checker, certifying the corresponding full-interval
bounds from Section~\ref{sec:qsp-verification}. At fixed
\(\epsilon_{\mathrm{QSP}}=10^{-4}\), the QSP degree increases from
3 to 77 as \(\tau\) increases from 0.1 to 50, while verification time
increases from 0.40 to 3.07 seconds. At fixed \(\tau=0.5\), all
circuits pass as the requested tolerance decreases from \(10^{-1}\)
to \(10^{-12}\). The largest degree-77 case spends 3.04 seconds in
symbolic execution and polynomial extraction, compared with
approximately 0.025 seconds in the final approximation check.
Symbolic processing therefore accounts for most of its runtime.
Peak RSS across the 50 runs is 67.02 MiB.

\paragraph{Composition of verified circuits.}
The block encodings verified in TASK1 can be composed
with the QSP wrappers verified in TASK2 following
Section~8.1.
Although TASK2 uses a single block-encoding ancilla,
its four-block abstraction extends to multiple ancillas by lifting the all-zero projector controls.

More precisely, suppose TASK1 certifies
$\|\alpha B_{\mathrm{BE}}-H\|\leq\epsilon_{\mathrm{BE}}$
and TASK2 certifies a polynomial $p$ satisfying
$\max_{x\in[-1,1]}|\beta p(x)-e^{-i\alpha tx}|
\leq\epsilon_{\mathrm{QSP}}$.
By the same triangle-inequality argument as in the proof of Theorem~8.1, the principal block of the composed circuit satisfies
\[
  \|\beta B_{\mathrm{comp}}-e^{-iHt}\|
  \leq\epsilon_{\mathrm{QSP}}
  +|t|\epsilon_{\mathrm{BE}}.
\]

For illustration, we combine the verified $n=256$
Ising block encoding in TASK1 with the largest-degree QSP
wrapper evaluated in TASK2 ($d=77$, $\tau=50$).
With $\epsilon_{\mathrm{BE}}=10^{-8}$,
$\epsilon_{\mathrm{QSP}}=10^{-4}$, and
$\alpha=512$, the composed circuit has
an error bound below $10^{-4}+\frac{50}{512}\cdot10^{-8} < 2\times10^{-4}$.
Its nominal macro-expanded size is
\[
  |C_{\mathrm{comp}}|
  =|C_{\mathrm{HS}}^{\mathrm{abs}}|-d+d|C_{\mathrm{BE}}|
  =2{,}333-77+77\cdot19{,}686
  =1{,}518{,}078
\]
gates, excluding the additional decomposition overhead for controlled
oracle calls and multi-controlled projector gadgets.
The verifier establishes this guarantee compositionally,
without expanding the block-encoding oracle calls
during QSP verification.

\subsection{Sensitivity to Injected Circuit Faults}
We inject representative circuit-construction faults into otherwise
accepted benchmarks. For the \(n=3\) Ising, MaxCut, and Heisenberg
block encodings, replacing a SELECT \(CZ\) gate with \(CX\) or
altering selector normalization produces \textsc{FAIL} in all 24
runs across four requested tolerances; all 12 corresponding clean
runs \textsc{PASS}. For four QSP wrappers with
\(\tau\in\{0.1,0.5,1,2\}\), moving an oracle call or omitting the
final Hadamard produces \textsc{FAIL} in all eight cases, while all
four clean wrappers \textsc{PASS}. Thus the verifier catches every
injected gate, normalization, and oracle-order fault in these tests.

We also perturb PREPARE angles and QSP phases. As
\(\epsilon_{\mathrm{BE}}\) tightens from \(10^{-1}\) to \(10^{-8}\),
the number of rejected block-encoding mutations rises from 6/15
to 15/15. At \(\epsilon_{\mathrm{QSP}}=10^{-4}\), all four
\(10^{-6}\) phase perturbations \textsc{PASS}, whereas all eight
larger perturbations \textsc{FAIL}. These results show that fault
detection is sensitive to the requested accuracy. As elsewhere in
this section, \textsc{FAIL} means that the checker could not certify
the bound.

\section{Related Works}
\paragraph{Hamiltonian-simulation and block-encoding compilers.}

Several recent systems provide high-level abstractions for constructing or
optimizing Hamiltonian-simulation and block-encoding circuits. 
QBlue~\cite{li2025verified} develops a verified compiler for Hamiltonian simulation, with
a circuit-generation pipeline centered on Trotterization-based simulation and
backend-specific synthesis.  In contrast, our work targets block-encoding and QSP-based simulation, and verifies an independently given circuit rather than a particular compiler pipeline. Cobble~\cite{yuan2026cobble} targets
quantum computational linear algebra and provides a high-level language for
programming with block-encoded matrices.  It compiles mathematical expressions
over block encodings into optimized quantum circuits.
SimuQ
~\cite{peng2024simuq} provides a framework for programming Hamiltonian simulation
with analog compilation.
Paulihedral~\cite{li2022paulihedral}, 2QAN~\cite{lao20222qan}, and MarQSim~\cite{cao2025marqsim} develop compilation and optimization techniques for
Hamiltonian-simulation workloads, often targeting Pauli-string-level
representations and hardware-efficient implementations. These systems are complementary to ours.  They focus on generating correct or
efficient circuits from high-level descriptions.  Our work instead addresses
compiler-independent verification: given a Hamiltonian specification and an
independently provided circuit, possibly produced by a compiler, optimizer, or
manual construction, our verifier checks whether the circuit satisfies the
intended block-encoding and QSP-level specification.

\paragraph{General-purpose quantum program verification.}
Quantum Hoare logic and its variants provide expressive proof systems for
reasoning about quantum programs~\cite{
ying2012floyd,zhou2019applied,unruh2019quantum,feng2021quantum,liu2019formal}.
These approaches can specify rich semantic properties using quantum predicates,
often represented by Hermitian operators or projections.  Approximate correctness has also been studied in Hoare-style frameworks, such
as~\cite{yan2024approximate}, which reasons about approximate relational properties
between two quantum programs.
However, they usually
require substantial user guidance and are commonly implemented in interactive
theorem provers such as Coq or Isabelle, or in deductive frameworks that still
require manual proof effort.

Existing automated circuit-analysis tools include symbolic verifiers based on
automata or SMT~\cite{ChenCLLTY23,ChenCLLT23,chen2025autoq,abdulla2026parameterized,chen2023theory}
and equivalence checkers based on ZX-calculus, path-sums, decision diagrams, or
stabilizer methods~\cite{
Coecke_2011,amy2018towards,burgholzer2020advanced,ChenJH22,WeiTJJ22,
xu2022quartz,ThanosCL23}.  These tools typically verify state-based
input-output properties or compare the implementation against another circuit.
Our setting is different: the specification is an operator-level Hamiltonian, and
the verifier checks that the given circuit satisfies the required block-encoding
and QSP correctness conditions and approximates the target evolution
$e^{-\i Ht}$ within a prescribed error bound, without requiring a reference
circuit for $e^{-\i Ht}$.

\bibliography{reference}

\newpage
\appendix

\section{Optimized QSP Circuit Construction for Hamiltonian Simulation}
\label{app:optimized-hs-construction}

This appendix describes the optimized QSP circuit construction used to generate
the TASK2 Hamiltonian-simulation benchmarks.  The purpose of this construction
is to produce a nontrivial optimized circuit-generation workflow whose output
can be independently checked by the verifier.  The construction itself is not
trusted: the verifier symbolically executes the emitted QASM circuit and checks
the polynomial implemented by the all-zero branch.

\paragraph{Cosine--sine decomposition.}

The Hamiltonian-simulation target on the normalized signal variable
\(x\in[-1,1]\) is
\[
    e^{-\mathrm{i}\tau x}
    =
    \cos(\tau x)-\mathrm{i}\sin(\tau x).
\]
The generator synthesizes the cosine and sine components separately.  This
matches the parity structure of standard QSP responses: an even-degree response
contains only even powers of \(x\), while an odd-degree response contains only
odd powers of \(x\).  Thus the cosine component is synthesized as an even
polynomial, and the sine component is synthesized as an odd polynomial.

For each effective simulation time \(\tau\), \textsc{pyqsp} is used only as an
untrusted source of benchmark phase sequences and polynomial data.  It produces
bounded Chebyshev approximations
\[
    \widehat P_{\cos}^{(\tau)}(x)
    \approx
    \frac{1}{2}\cos(\tau x),
    \qquad
    \widehat P_{\sin}^{(\tau)}(x)
    \approx
    \frac{1}{2}\sin(\tau x).
\]
The factor \(1/2\) is the bounded-synthesis scaling used in the generated
benchmark data.  It is accounted for in the final normalization step.

\paragraph{Sharing one QSP skeleton.}

A direct implementation would emit two independent QSP circuits: one for
\(\widehat P_{\cos}^{(\tau)}\) and one for
\(\widehat P_{\sin}^{(\tau)}\).  This is simple but duplicates the alternating
sequence of \(Q_{\mathrm{BE}}\) and \(Q_{\mathrm{BE}}^\dagger\) calls.  The
optimized construction avoids this duplication by using a selector qubit to
multiplex the two phase sequences over a shared QSP skeleton.

Let \(\psi_j^{\cos}\) and \(\psi_j^{\sin}\) denote the projector phases used at
layer \(j\) for the cosine and sine QSP sequences, respectively.  The circuit
applies the cosine projector phase \(\psi_j^{\cos}\) by default.  On the sine
branch, it applies the controlled phase difference
\[
    \Delta_j=\psi_j^{\sin}-\psi_j^{\cos}.
\]
Therefore the effective phase is \(\psi_j^{\cos}\) on the cosine branch and
\(\psi_j^{\cos}+\Delta_j=\psi_j^{\sin}\) on the sine branch.  Hence the two
selector branches implement the cosine and sine QSP sequences while sharing the
same positions of \(Q_{\mathrm{BE}}\) and \(Q_{\mathrm{BE}}^\dagger\) calls.  If
the sine degree is one larger than the cosine degree, the generator appends one
final sine-only block on the sine branch.

\paragraph{Phase-convention conversion.}

The phase sequence returned by \textsc{pyqsp} cannot be inserted directly as
QASM \(R_z\) rotations, because \textsc{pyqsp} and our block-encoding QSVT
circuit use different scalar signal conventions.  In the \textsc{pyqsp}
convention, the scalar signal unitary is
\[
    W_x(x)
    =
    \begin{pmatrix}
        x & \mathrm{i}\sqrt{1-x^2} \\
        \mathrm{i}\sqrt{1-x^2} & x
    \end{pmatrix}.
\]
Our block-encoding QSVT circuit instead corresponds to
\[
    W'_x(x)
    =
    \begin{pmatrix}
        x & \sqrt{1-x^2} \\
        \sqrt{1-x^2} & -x
    \end{pmatrix}.
\]
These conventions differ only by fixed \(Z\)-axis phase shifts.  The generator
absorbs these fixed shifts into the neighboring projector phases before emitting
QASM.

Given \textsc{pyqsp} phases \(\phi_0,\ldots,\phi_d\), the generator converts
them to projector phases by setting \(\psi_0=\phi_0+\pi/4\) and
\(\psi_j=\phi_j+\pi/2\) for \(1\le j\le d-1\).  The final phase depends on the
component:
\[
    \psi_d=\phi_d+\frac{\pi}{4}
    \quad
    \text{for the even cosine component,}
    \qquad
    \psi_d=\phi_d-\frac{\pi}{4}
    \quad
    \text{for the odd sine component.}
\]
A projector phase \(\psi_j\) is emitted in QASM as \(R_z(-2\psi_j)\).  This
conversion is part of the benchmark generator only; the verifier does not
assume that it is correct.

\paragraph{Extracting the polynomial response.}

Under the above convention, the selected raw QSP response contains the desired
polynomial in its imaginary part.  For \(\star\in\{\cos,\sin\}\), the raw
selected branch has the form
\[
    B_\star^{\mathrm{raw}}(x)
    =
    R_\star(x)
    +
    \mathrm{i}\widehat P_\star^{(\tau)}(x),
\]
where \(R_\star(x)\) denotes the remaining real component.  The polynomial that
should be combined by the outer LCU is \(\widehat P_\star^{(\tau)}(x)\), not the
whole raw response.  The generator therefore inserts a constant-size signed
phase gadget that forms
\[
    \frac{
        B_\star^{\mathrm{raw}}(x)
        -
        B_\star^{\mathrm{raw}}(x)^\dagger
    }
    {2\mathrm{i}}
    =
    \widehat P_\star^{(\tau)}(x).
\]
After this step, the cosine branch exposes
\(\widehat P_{\cos}^{(\tau)}(x)\), and the sine branch exposes
\(\widehat P_{\sin}^{(\tau)}(x)\).

\paragraph{LCU combination and normalization.}

The extracted cosine and sine polynomials are combined by a two-term LCU.  The
outer selector prepares an equal superposition of the cosine and sine branches,
assigns the coefficient \(-\mathrm{i}\) to the sine branch, and then unprepares
the selector.  Projecting back to the all-zero selector branch yields
\[
    B^{\mathrm{HS}}_{\mathbf 0}(x)
    =
    \frac{1}{2}
    \left(
        \widehat P_{\cos}^{(\tau)}(x)
        -
        \mathrm{i}\widehat P_{\sin}^{(\tau)}(x)
    \right).
\]
The prefactor \(1/2\) is the success amplitude of the two-term LCU: one factor
\(1/\sqrt{2}\) comes from preparing the selector superposition, and another
factor \(1/\sqrt{2}\) comes from unpreparing it.

Since the synthesized QSP polynomials are themselves half-scaled,
\[
    \widehat P_{\cos}^{(\tau)}(x)
    \approx
    \frac{1}{2}\cos(\tau x),
    \qquad
    \widehat P_{\sin}^{(\tau)}(x)
    \approx
    \frac{1}{2}\sin(\tau x),
\]
the all-zero branch satisfies
\[
    B^{\mathrm{HS}}_{\mathbf 0}(x)
    \approx
    \frac{1}{2}
    \left(
        \frac{1}{2}\cos(\tau x)
        -
        \mathrm{i}\frac{1}{2}\sin(\tau x)
    \right)
    =
    \frac{1}{4}e^{-\mathrm{i}\tau x}.
\]
Thus the generated circuit implements the desired Hamiltonian-simulation
function up to the known scaling factor \(1/4\).  The verifier infers a scaling factor \(\beta\) from the extracted branch.
For these benchmarks, the inferred value is close to \(4\), as expected
from the construction, and the verifier checks
\[
    \max_{x\in[-1,1]}
    \left|
        \beta B^{\mathrm{HS}}_{\mathbf 0}(x)
        -
        e^{-\mathrm{i}\tau x}
    \right|
    \le
    \epsilon .
\]
\end{document}